%% file: main_ieee.tex
\documentclass[journal]{IEEEtran}
\IEEEoverridecommandlockouts
\usepackage{cite}
\usepackage{amsfonts}
\usepackage{textcomp}
\usepackage[table,usenames,dvipsnames]{xcolor}
\usepackage{hyperref}
\usepackage{float} 
\usepackage{subcaption}
\usepackage{algorithm}
\usepackage{algcompatible}
\usepackage[noend]{algpseudocode}
\usepackage{soul}
\usepackage{subfiles}
\usepackage{multirow}
\usepackage{hhline}
\usepackage{xpatch}
\usepackage{balance}
\usepackage{placeins}
\usepackage{tikz}
\usepackage{bm}
\usepackage{boldline}
\usepackage{authblk}
\usepackage{orcidlink}
\usepackage{amsmath}
\usepackage{booktabs}
\usepackage{tabularx}

\usepackage{graphicx}
\usepackage{amssymb}
\usepackage{mathtools}
\usepackage{physics}
\usepackage{overpic}
\usepackage{array}
\usepackage{fancyvrb}
\newcolumntype{C}{>{\centering\arraybackslash}X}

\makeatletter
\def\thanks#1{\protected@xdef\@thanks{\@thanks
        \protect\footnotetext{#1}}}
\makeatother

\definecolor{turquoise}{cmyk}{0.65,0,0.1,0.3}
\definecolor{purple}{rgb}{0.65,0,0.65}
\definecolor{dark_green}{rgb}{0, 0.5, 0}
\definecolor{orange}{rgb}{0.8, 0.6, 0.2}
\definecolor{red}{rgb}{0.8, 0.2, 0.2}
\definecolor{darkred}{rgb}{0.6, 0.1, 0.05}
\definecolor{blueish}{rgb}{0.0, 0.3, .6}
\definecolor{light_gray}{rgb}{0.7, 0.7, .7}
\definecolor{pink}{rgb}{1, 0, 1}
\definecolor{greyblue}{rgb}{0.25, 0.25, 1}
\definecolor{blue_bar}{RGB}{5,50,255}
\definecolor{red_bar}{RGB}{251,1,4}
\definecolor{green_bar}{RGB}{34,255,5}
\definecolor{pink_bar}{RGB}{251,3,255}

\DeclareMathOperator*{\argmin}{arg\,min}

\usepackage{blindtext}

\renewcommand{\paragraph}[1]{\vspace{1em}\noindent\textbf{#1}.}

\usepackage{bbding}
\usepackage{pifont}
\usepackage{wasysym}

\usepackage{tikz}
\definecolor{tagbg}{RGB}{225,236,244}
\definecolor{helabg}{RGB}{236,236,244}
\definecolor{helaabg}{RGB}{210,236,230}
\definecolor{bactag}{RGB}{190,236,190}

\definecolor{mnist_txt}{RGB}{200,226,255}
\definecolor{mnist2_txt}{RGB}{244,193,66}
\definecolor{hela_txt}{RGB}{255,180,190 }
\definecolor{helaa_txt}{RGB}{212,255,212}
\definecolor{bac_txt}{RGB}{251,212,255}

\definecolor{tagtxt}{RGB}{0,0,0}

\newcommand{\journalname}{} 

\def\BibTeX{{\rm B\kern-.05em{\sc i\kern-.025em b}\kern-.08em
    T\kern-.1667em\lower.7ex\hbox{E}\kern-.125emX}}
\makeatletter
\xpatchcmd{\algorithmic}{\itemsep\z@}{\itemsep=.03cm}{}{}
\makeatother

\usepackage{xspace}

\makeatletter
\DeclareRobustCommand\onedot{\futurelet\@let@token\@onedot}
\def\@onedot{\ifx\@let@token.\else.\null\fi\xspace}

\def\eg{\emph{e.g}\onedot}

\def\etal{\emph{et al}\onedot}
\makeatother

\title{Differentiable Microscopy Designs an All Optical Phase Retrieval Microscope}
\author{Kithmini Herath$^{1, 2}$, Hasindu Kariyawasam$^{\dagger1, 2}$,  Ramith Hettiarachchi$^{\dagger 1, 2}$, Udith Haputhanthri$^{\dagger 1, 2}$, Dineth Jayakody$^{3}$, Raja Naeem Ahmad$^{1}$, Azeem Ahmad$^{4}$, Balpreet S. Ahluwalia$^{4}$, Chamira U. S. Edussooriya$^{2}$\IEEEauthorrefmark{3} and Dushan N. Wadduwage$^{1, 3, 5}$\IEEEauthorrefmark{1}
\thanks{$^\dagger$ These authors contributed equally.}
\thanks{$^{1}$Center for Advanced Imaging, Faculty of Arts and Sciences,  Harvard University, Cambridge, MA 02138, USA}%
\thanks{$^{2}$Department of Electronic and Telecommunication Engineering, University of Moratuwa, Sri Lanka}%
\thanks{$^{3}$Department of Computer Science, Old Dominion University, Norfolk, VA, USA}%
\thanks{$^{4}$Department of Physics and Technology, UiT The Arctic University of Norway, Tromsø 9037, Norway}%
\thanks{$^{5}$Department of Physics and School of Data Science, Old Dominion University, Norfolk, VA, USA \\}%
\thanks{\textit{Emails: \IEEEauthorrefmark{1}dwadduwa@odu.edu, \IEEEauthorrefmark{3}chamira@uom.lk}}}

\begin{document}
\maketitle

\begin{abstract}
Designing new optical systems from the ground up for microscopy imaging tasks {\color{black}such as phase
retrieval,}   requires substantial scientific expertise and creativity. {\color{black} To augment the traditional design process}, we propose differentiable microscopy ($\partial\mu$), which introduces a top-down design {\color{black} approach}. Using all-optical phase retrieval as an illustrative example, we demonstrate the effectiveness of data-driven microscopy design through $\partial\mu$. Furthermore, we conduct comprehensive comparisons with {\color{black}existing all-optical phase retrieval} methods, showcasing the consistent superiority of our learned designs across multiple datasets, including biological samples. To substantiate our ideas, we experimentally validate the functionality of one of the learned designs, providing a proof of concept. The proposed differentiable microscopy framework supplements the creative process of designing {\color{black}new phase microscopy systems and may be extended to other similar applications in optical design.}  
\end{abstract}


\input{sec/1_introduction}

\input{sec/2_related_work}
\input{sec/3_preliminaries}
\input{sec/4_methodology}

\input{sec/5_results}
\input{sec/6_conclusions}

\section*{Acknowledgement}
The authors would like to acknowledge Dr. Quansan Yang, Dr. Gaojie Yang, and Prof. Edward S. Boyden of Massachusetts Institute of Technology, Cambridge, USA for the fabrication of the Phase MNIST samples used for the experimental validation.

\balance
\bibliographystyle{IEEEtran}
\bibliography{macros,refs,references,references2}
\setcounter{section}{0}
\setcounter{figure}{0}
\setcounter{table}{0}
\renewcommand{\thesection}{\Roman{section}}
\renewcommand{\thefigure}{S\arabic{figure}}
\renewcommand{\thetable}{S\arabic{table}}
\input{sec/X_supplementary_v2}


\end{document}

%% file: sec/1_introduction.tex
\section{Introduction}
\label{sec:intro}
\input{fig/teaser}

Measuring phase information of a light field is a longstanding problem in optics with valuable applications such as live cell imaging \cite{Majeed2017QuantitativeDiagnosis,Mir2010BloodCytometry,Roitshtain2017,Marquet2014ReviewDisorders,Hu2019}. Objects like cells are thin and transparent. They are referred to as phase objects as they modulate the phase of the light field, not the intensity. Imaging devices however only measure the intensity of light.  To measure the phase, interferometric systems, called quantitative phase microscopes (QPMs), are needed. QPMs interfere the phase object's light wave (object wave) with a second known light wave (called the reference wave) resulting in interference patterns (called interferograms). Phase images are then reconstructed from  interferograms, by computationally solving an inverse problem. {\color{black} Designing such an instrument} involves understanding: the light-matter interaction at the specimen, how light propagates through optics, and how the final image is formed. The optical scientist’s job is to design the optics such that the final image contains desired information about the measurands. This creative process may require enormous fuzzy calculations on the light propagation just to initialize a potential optical design. In this work, we present a data-driven top-down approach to supplement this creative design process {\color{black} and demonstrate it for the task of phase microscopy}.


Data-driven machine learning approaches have been increasingly used in various microscopy methods to perform tasks such as denoising, image reconstruction, and classification \cite{Viz2021, Park2021, Cha2025, Xue2019ReliableQuantification, Wang2019DeepMicroscopy, Moh2022}. Despite the significant performance boost in such methods, the limitations inherited from the hardware of the microscope set the upper performance bound \cite{Cooke2021Physics-enhancedMicroscopy}. Therefore, over the past decade, researchers have focused on joint data-driven optimization of not only the reconstruction model but also the hardware of the microscope itself \cite{Chakrabarti2016LearningBack-propagation, Sun2020LearningSelection, Chaware2020TowardsClassification, Muthumbi2019LearnedClassification, Horstmeyer2017ConvolutionalImage, Kim2020Multi-elementLayers,haim2018depth, elmalem2018learned, wu2019phasecam3d}. 
{\color{black}Building on these foundations,} we introduce \emph{differentiable microscopy} ($\partial \mu$) as a top-down design approach where new designs can be parameterized and optimized first as black-box models and then as optical architectures (see Fig.~\ref{fig:teaser}). 
{\color{black}As a first step, in this paper, we introduce and demonstrate the utility of this approach for the task of all-optical phase retrieval (see Fig.~\ref{fig:teaser}).}

 Unlike QPM, all-optical phase retrieval does not require computational reconstruction. The interference pattern (i.e. the  intensity recorded on the camera) itself should be the phase image. All-optical phase retrieval is a challenging design task with no universal solution appropriate for all possible phase objects. The first approach, i.e. phase contrast, was introduced by Zernike~\cite{Zernike1942PhaseObjects,Zernike1942PhaseII}. Following Zernike, Gl\"{u}ckstad~\cite{Gluckstad1996} introduced the generalized phase contrast (GPC) method. Both these were bottom-up designed, starting from design rules to generate desired interferences through mathematical treatment of the problem. More recently, in parallel to our work \cite{herath2022_old}, Mengu \textit{et al.} \cite{ozcan_phase} demonstrated all-optical phase retrieval using a diffractive deep neural network (D2NN).  This is a top-down design approach, and they trained and analyzed the D2NNs on a number of datasets.  Our work differs from Mengu \textit{et al.} \cite{ozcan_phase}, in the way we initialize the top-down design using inductive biases, interpretability of learned designs, the use of data with complex biological features, and experimental demonstration in the visible wavelengths. Importantly, we replicated the D2NN, and GPC methods, to compare and contrast all three (i.e. D2NN, GPC, and ours) in an unbiased fashion on the same datasets in numerical experiments. 

In brief, our main contributions are as follows:

\begin{enumerate}
    \item We introduce our proposed Differentiable Microscopy ($\partial \mu$) approach, and {\color{black} investigate its utility across } three  {\color{black}distinctive} optical architectures designed for all-optical phase retrieval.

    \item We evaluate the performance of the Generalized Phase Contrast (GPC) method for the same purpose. To gauge the effectiveness, we numerically test multiple variants of all the approaches, on multiple datasets. Further, we critically review all designs yielding intriguing insights into the top-down design of microscopes.
    \item {\color{black} We provide a proof-of-concept experimental validation for a learned Fourier-plane filter design.}
\end{enumerate}

{\color{black} To avoid potential confusion, we note that, while $\partial\mu$ offers a top-down design approach, the learned microscopes remain limited by the underlying physics of the forward model (see Section~\ref{sec:discussion_assumptions_scope} for a detailed discussion).}

The remaining of the paper is organized as follows. In Sec.~\ref{sec:related_work}, we review related works to our study. We briefly review optical filters in the Fourier plane and D2NNs in Sec~\ref{sec:preliminaries}. In Sec.~\ref{sec:methodology}, we present $\partial \mu$ and three top-down approaches employed for phase-to-intensity conversion. In Sec.~\ref{sec:results}, we present experimental results and a discussion. Finally, conclusion is presented in Sec.~\ref{sec:con}.

%% file: fig/teaser.tex
\begin{figure}[t]
\begin{center}
\includegraphics[width=\linewidth]{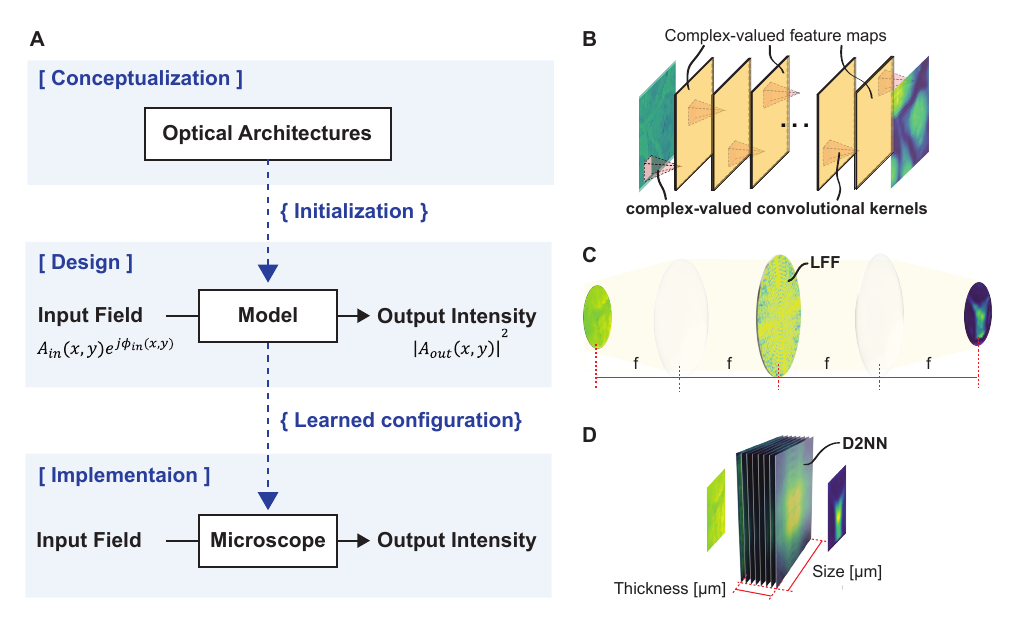}
\end{center}
\caption{
\textbf{Differentiable microscopy ($\partial \mu$)  based all-optical phase to intensity conversion.} \textbf{(A)} Problem formulation as a learning task and model implementation pipeline in differentiable microscopy. \textbf{(B)} The complex-valued linear CNN architecture that was implemented mimicking optical constraints for linear phase retrieval. Note that there are no non-linear activations between convolutions. \textbf{(C)} Learnable Fourier filter (LFF) based design. Here, $f$ denotes the focal length of the lens used to implement the 4-\textit{f} system. \textbf{(D)} Diffractive deep neural network based design (D2NN). Note that D2NN is very compact and less than $50$ $\mu m$ (i.e. $51.7\lambda$, where $\lambda$ is the operating wavelength) thick.}
\label{fig:teaser}
\end{figure}

%% file: sec/2_related_work.tex
\section{Related Work}
\label{sec:related_work}


\subsection{Optical Hardware Optimization}
Rather than following the static optical setup of the microscope and post-processing its acquired images, recent methods have focused on optimizing certain parts of the optical hardware itself. Several approaches focus on optimizing the illumination patterns of the microscope \cite{Chaware2020TowardsClassification,Muthumbi2019LearnedClassification,Horstmeyer2017ConvolutionalImage,Kim2020Multi-elementLayers}. This research direction of jointly optimizing the forward optics (by learning illumination patterns) with the inverse reconstruction model has been able to reduce the data requirement in QPM \cite{Kellman2019Data-DrivenMicroscopy,Kellman2019Physics-BasedImaging}. {\color{black} Beyond illumination, there is a growing body of work in end-to-end differentiable design of phase masks for hybrid sensing systems \cite{haim2018depth, elmalem2018learned, wu2019phasecam3d}. Our $\partial\mu$ approach builds on these foundations but focuses on the discovery of all-optical architectures where the physical system itself performs the final reconstruction.}
This emerging field of automated design operates at different levels of abstraction. For instance, the XLuminA framework by Rodríguez et al. \cite{rodriguez2024automated} tackles the discovery of both the optical topology and its parameters, and discovers novel layouts and physical concepts for super-resolution microscopy. Our $\partial\mu$ framework approaches the design process itself by learning novel hardware configurations within a given class of optical architecture, such as a $4$-$f$ system or a D2NN. This methodology allows us to discover new, data-driven designs and their interpretable rules in the context of well-understood and physically realizable frameworks, which we demonstrate for the task of all-optical phase retrieval.



\subsection{Fourier Plane Filter Optimization}
The 4-\textit{f} optical system (Fig.~\ref{fig:teaser}C) produces the Fourier transform of the incident spatial light at the Fourier plane~\cite{goodman2005}. This Fourier transform property of a lens can be leveraged to perform the convolution operation through light itself by placing filters at the Fourier plane. The effectiveness of this optical convolution has been studied in the literature to perform various classification tasks~\cite{Chang2018HybridClassification,Burgos2021DesignNetworks,Colburn2019OpticalNetwork,Wu2021Multi-layerTheorem}.

\subsection{Diffractive Deep Neural Networks}
D2NNs are a type of optical neural networks which has been introduced by Lin \etal.~\cite{Lin2018} as a physical mechanism to perform machine learning. They consist of a set of diffractive surfaces acting as the layers of the machine learning model. The transmission coefficients of each diffractive element act as the neurons of the network and they can be trained by modeling them with optical diffraction theory~\cite{Lin2018,Arg2023}. The trained diffraction surfaces can be 3D printed and inference is done by propagating light through the D2NN. Since they work purely optically, the D2NNs has the capability of performing computations at the speed of light with no power consumption. D2NNs are utilized for various tasks such as image classification~\cite{Li2018, Li2019Class-specificAccuracy}, object detection and segmentation~\cite{Yan2019Fourier-spaceNetwork}, designing task-specific optical systems~\cite{Luo2019DesignNetworks}, reconstruction of overlapping phase objects~\cite{Mengu2021}, and image reconstruction using holograms~\cite{SakibRahman2021Computer-FreeNetworks} and quantitative phase imaging \cite{ozcan_phase}. {\color{black}Moreover, unlike Fourier-space D2NNs~\cite{Yan2019Fourier-spaceNetwork} which utilize multi-layer volumetric processing for semantic tasks, our work investigates the sufficiency of single-layer planar filtration for phase retrieval.}

\subsection{All-Optical Phase Retrieval Methods}
The work of Zernike~\cite{Zernike1942PhaseObjects,Zernike1942PhaseII} paved the way to clearly visualize transparent biological samples under a microscope without applying contrast dyes onto the specimen. Thus, the development of phase contrast microscopy is a significant milestone in biological imaging. This phase contrast method is implemented in a common-path interferometer where the signal and reference beams resulting from the illumination travel along the same optical axis and interfere at the output of the optical system generating an interferogram. The phase perturbations introduced by the optical characteristics (refractive index, thickness) of the object are converted to observable intensity variations on the interferogram, by the use of a Fourier plane filter in the optical path. This filter functions as a phase shifting filter which imparts a quarter-wave shift on the undiffracted light components (the reference beam). However, the linear relationship between the input phase and output intensity of this system is derived based on the ``small-scale" phase approximation, where the largest input phase deviation is considered to be $\pi/3$. Gl\"{u}ckstad~\cite{Gluckstad1996} generalized Zernike's phase contrast method and introduced the generalized phase contrast (GPC) method to provide an all-optical linear phase to intensity conversion approach, which overcomes the restriction of ``small-scale" phase approximation. The descriptive mathematical analysis of the filter implementation in \cite{Gluckstad2009} shows the utilization of a phase shifting Fourier plane filter, where its parameters are carefully selected to appropriately generate the synthetic reference wave to reveal the input phase perturbations in the output interference pattern. Through further mathematical analysis they have demonstrated how the system parameters can be optimized to produce accurate phase variation measurements and reduce the error in quantitative phase microscopy when a GPC-based method is utilized with post-processing. Even though the GPC method is capable of linearly mapping the input phase information to the output intensity all optically, it requires a careful mathematical analysis, a considerable amount of domain knowledge and relatively complex optical setups with post-processing to implement quantitative phase imaging. On the other hand, in our implementation the microscope is optimized for phase reconstruction as an end-to-end differentiable model. Furthermore, recently, Mengu \textit{et al.} \cite{ozcan_phase} trained D2NNs on datasets such as MNIST, CIFAR10, TinyImageNet, FashionMNIST and demonstrated the all-optical phase to intensity conversion task. They tested a trained network on Pap-smear samples (monolayer of cells). It is also important to notice that the training strategy (e.g., the loss function) employed in this work is different from the phaseD2NN demonstrated in our work. {\color{black} Moreover, Shen \textit{et al.}~\cite{ShenPhaseD2NN2023} extended diffractive processing to multispectral quantitative phase imaging, utilizing a deep multi-layer D2NN. In contrast to their deep network approach, our work investigates the efficacy and interpretability of single-layer learnable filters while quantifying the performance of D2NNs on phase retrieval.} 

%% file: sec/3_preliminaries.tex
\section{Preliminaries}
\label{sec:preliminaries}

\subsection{Optical Filters in the Fourier Plane}
\begin{figure}[t!]
\centering
\includegraphics[width=\linewidth]{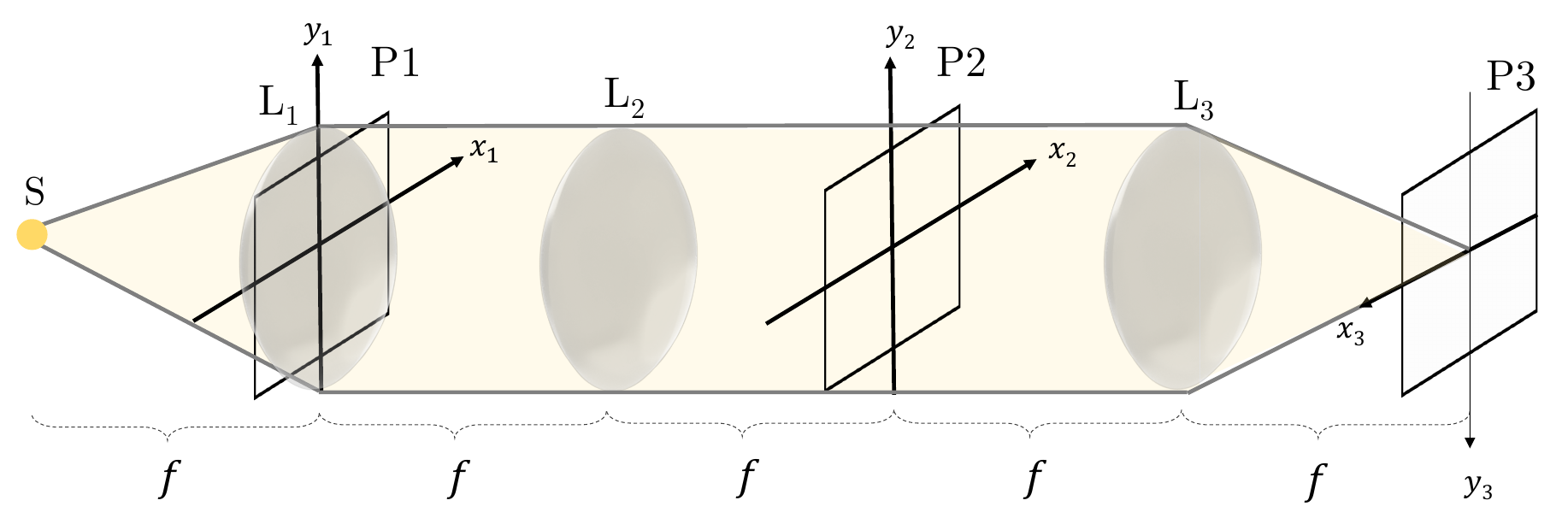}
\caption{4-\textit{f} system, where the distance between the input plane ($P_1$) and the image plane ($P_3$) equals to $4 \times f$. Here, $L_1, L_2, L_3$ are identical lenses each having a focal length $f$, $P_1, P_2, P_3$ are the input plane, Fourier plane and the image plane, respectively, and $S$ is a point source.} 
\label{fig:4f}
\end{figure}

We first review the 4-\textit{f} system~\cite{goodman2005}, shown in Fig. \ref{fig:4f}, that can be employed to implement an optical filter. This system is referred as 4-\textit{f} because of the distance between the input plane ($P_1$) and the image plane ($P_3$) equals to $4 \times f$, where $f$ is the focal length of the lenses. Here $S$ is a point source, $P_1, P_2, P_3$ are the input plane, Fourier plane and the image plane, respectively, and $L_1, L_2, L_3$ are identical lenses each having a focal length $f$. The light from $S$ is collimated with the lens $L_1$. The input sample ($P_1$) which has $g(x_1, y_1)$ amplitude transmission coefficient is placed against $L_1$ to reduce the overall length of the system. The resultant light is then collimated with the Fourier transforming lens $L_2$ resulting $k_1G(\frac{x_2}{\lambda f}, \frac{y_2}{\lambda f})$ field at $P_2$ plane where $G$ is the Fourier transform of $g$ and $k_1$ is a constant. To manipulate the spectrum of input field $g(x_1, y_1)$, filter can be placed in the Fourier plane in $P_2$. The transmission coefficient of the filter is given by 
\begin{equation}
t_{A}\left(x_{2}, y_{2}\right)=k_{2} H\left(\frac{x_{2}}{\lambda f}, \frac{y_{2}}{\lambda f}\right),
\label{eq:4f_transmission}
\end{equation}
where the desired frequency domain transfer function is $H$ and $k_2$ is a constant. Following lens $L_3$ applies the Fourier transform again on the modified spectrum to obtain the inverted modified field. The GPC introduced by Gl\"{u}ckstad  \cite{Gluckstad2009} is a circular filter in the Fourier plane consisting of 2 main regions, namely; the central region and the outer region. In this implementation, a phase shift is introduced by the central region of the filter which results in an interference pattern that produces a substantial contrastive image of the input phase object. 

\subsection{Diffractive Deep Neural Networks}
\label{sec:prelim_d2nn}
We next review the D2NNs and modeling of the light propagation in a D2NN. A D2NN comprises of a set of diffractive layers. Each diffractive element (neuron) in a diffractive layer can be considered as a secondary wave source according to the Huygens-Fresnel principle~\cite{Lin2018}. The diffraction at each neuron can be modelled using the Rayleigh-Sommerfeld diffraction formulation~\cite[ch. 3.5]{goodman2005} as follows. 

\begin{figure}[t!]
\centering
\includegraphics[width=6.5cm, height=5cm]{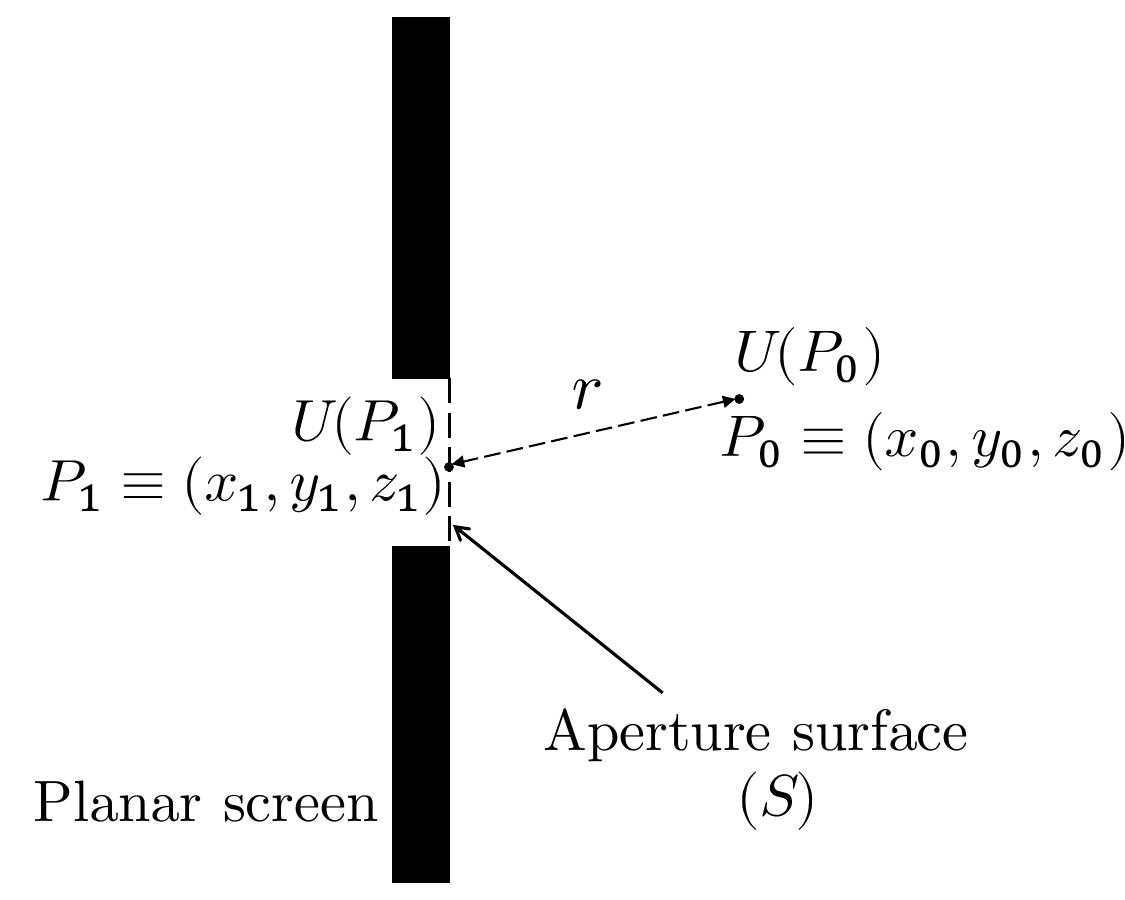}
\caption{Diffraction by an aperture in a planar screen.}
\label{fig:diffraction_rs}
\end{figure}

The Rayleigh-Sommerfeld diffraction theory describes the diffraction of light by an aperture in an infinite opaque planar screen. Consider such an aperture illuminated by a light wave which has a field $U(P_1)$ at a point $P_1 \equiv (x_1,y_1,z_1)$ on the aperture surface $S$ as shown in Fig.~\ref{fig:diffraction_rs}. The field at point $P_0 \equiv (x_0,y_0,z_0)$ after the diffraction can be given using the first Rayleigh-Sommerfeld solution to the diffraction problem, which is given as
\begin{equation}
    U(P_0) = \iint\limits_S U(P_1)\,\left(\frac{z_0 - z_1}{r^2}\right) \left(\frac{1}{2\pi r} + \frac{1}{j \lambda}\right) e^{j \frac{2\pi r}{\lambda}}\,dS.
    \label{eq:rs1}
\end{equation}
Here, $r = \sqrt{(x_0 - x_1)^2 + (y_0 - y_1)^2 + (z_0 - z_1)^2}$ is the distance between the points $P_1$ and $P_0$, $\lambda$ is the wavelength of the light wave and $j = \sqrt{-1}$. This integral holds under the conditions that, $U$ is a homogeneous scalar wave equation and it satisfies the \emph{Sommerfeld radiation condition} which states that $U$ must vanish at least as fast as a diverging spherical wave~\cite[ch. 3.4]{goodman2005}. The integral in~\eqref{eq:rs1} can be directly converted to a Riemann summation considering a discrete set of sampling points on the aperture plane. In this direct implementation, each neuron in a diffractive layer is considered as a sampling point.

\begin{figure}[t!]
\centering
\includegraphics[width=0.9\linewidth]{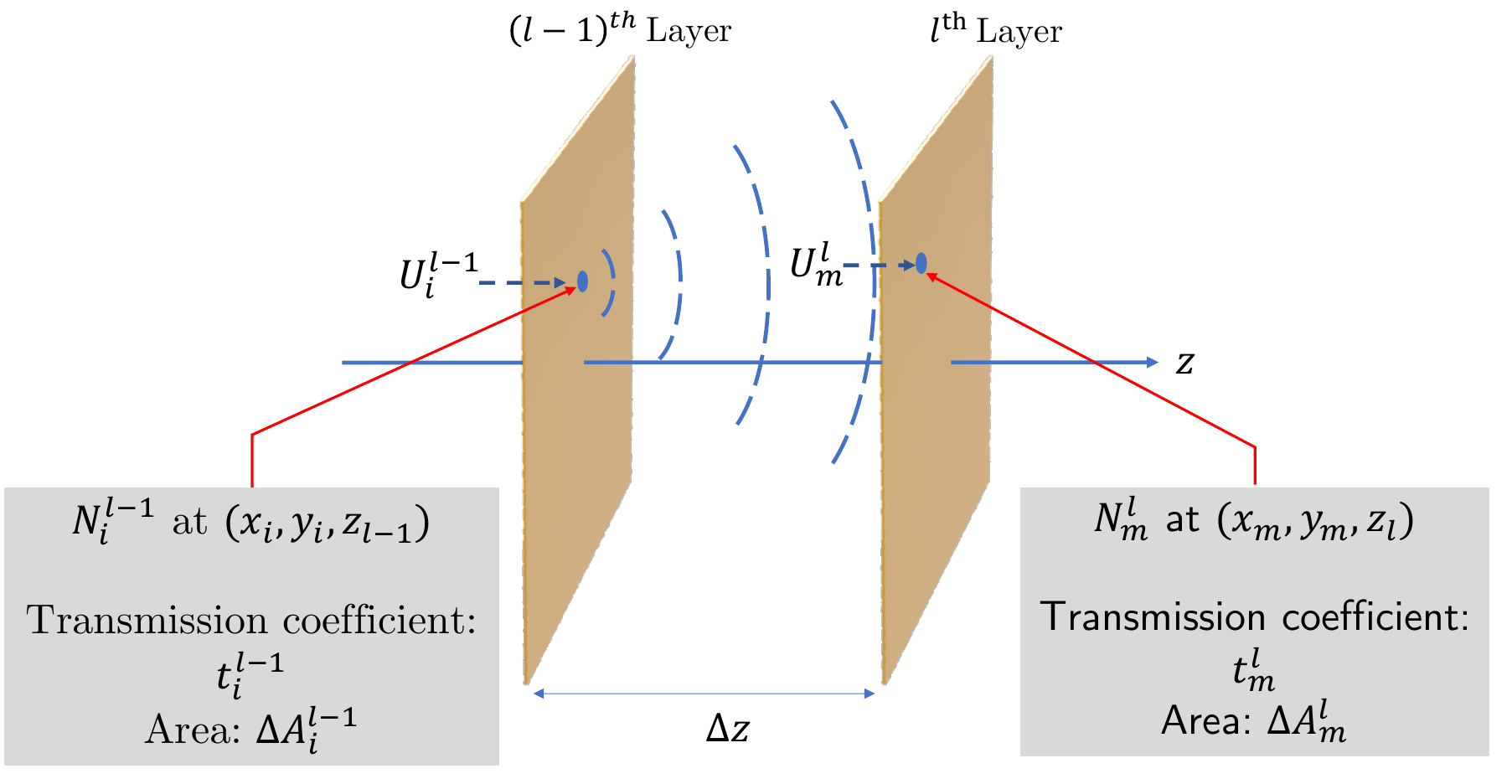}
\caption{Diffraction of light waves by a neuron of a D2NN layer.}
\label{fig:diffractive_layer}
\end{figure}

As shown in Fig.~\ref{fig:diffractive_layer}, suppose that the field at the $i^{\textrm{th}}$ neuron of the $(l-1)^{\textrm{th}}$ layer ($N_{i}^{l-1}$), which is located at $(x_i, y_i, z_{l-1})$ is given by $U_{i}^{l-1}$, and the diffraction caused at this neuron results in the field given by $U_{m,i}^{l}$ at $N_{m}^{l}$ located at $(x_m, y_m, z_l)$. Then, $U_{m,i}^{l}$ is given by,
\begin{equation}
    U_{m,i}^{l} = U_{i}^{l-1}\,t_{i}^{l-1}\,w_{m,i}^{l}\,\Delta A_{i}^{l-1},
\end{equation}
where
\begin{equation}
    w_{m,i}^{l} = \left(\frac{\Delta z}{r_{m,i}^2}\right) \left(\frac{1}{2\pi r_{m,i}} + \frac{1}{j \lambda} \right) \exp{\left(\frac{j2\pi r_{m,i}}{\lambda}\right)},
\end{equation}
where, $\Delta A_{i}^{l-1}$ is the area of $N_{i}^{l-1}$, $t_{i}^{l-1}$ is the transmission coefficient of the same neuron, $\Delta z = z_l - z_{l-1}$ is the distance between the two adjacent layers along the direction of light propagation, and $r_{m,i} = \sqrt{(x_m - x_i)^2 + (y_m - y_i)^2 + \Delta z^2}$. The overall resulting field at $N_{m}^{l}$ is the superposition of all the fields resulted by the diffraction at each of the neurons at the $(l-1)^{\textrm{th}}$ layer which is given by
\begin{equation}
    U_{m}^{l} = \sum_{i} U_{m,i}^{l}.
\end{equation}

We note that the direct implementation of these equations results in resource intensive computations in the simulations. A more computationally efficient modeling is done using the fact that the field resulted by an optical wave and its angular spectrum are related through the Fourier transform~\cite{Ratcliffe1956, goodman2005}. This method is known as the angular spectrum (AS) method and it is used for simulating D2NNs in our work. Further information about the AS method is described in the supplementary materials.

%% file: sec/4_methodology.tex
\section{Differentiable Microscopy: Proposed Learnable Filter and D2NN for All-Optical Phase Retrieval}
\label{sec:methodology}
We present $\partial \mu$ approach in detail and three top-down designs: a complex-valued CNN, a learnable Fourier filter and a D2NN for all-optical phase retrieval in this section. We first present the $\partial \mu$ approach in the next subsection.

\subsection{Differentiable Microscopy: Designing Optical Microscopes Top-down}
\label{sec:problem_formulation}
Classical optical system design is a bottom-up approach where physicists assemble optical elements in a certain configuration to manipulate light passing through them. In our example of all-optical phase retrieval, we know that the signal is in the phase of the light field, and hence we need to assemble optics to measure the phase of light. The phase of light is measured by interfering the object-light-field with a reference field. But, when this phenomenon is not known a priori we cannot initiate an optical design bottom-up. 

In $\partial \mu$, we present a top-down design approach to bypass this initialization problem. We start with the application, i.e., the desired input and output of the optical system. At this stage, the instrument design is thought of as a blackbox that translates the input light field to an output intensity field (see the design stage in Fig.~\ref{fig:teaser}A). Our goal is to assume a reasonable functional form (for the design) with a set of parameters, assuming at least one configuration of the parameters will approximate a function that can translate inputs to the desired outputs. When the functional form is known we can “train” the parameters by some supervised machine learning approach given paired input and outputs. We will later discuss how the input-output pairs can be formulated. We first discuss how a “reasonable functional form” can be designed. 

We know some constraints of the functional form. We know that it should be complex-valued, but linear (assuming no non-linear optical elements are used in the design). We know that it should be translational invariant laterally so that it is not sensitive to the position of objects in the sample. We first start with a function with these properties; in our example, we used a linear complex-valued convolutional neural network (see Fig.~\ref{fig:teaser}B given the empirical ease of optimizing such a function. We then transformed this design into an optical design as described below. Our CNN is a fully convolutional function with only linear layers. Thus, the final form of the CNN should be equivalent to a single convolution that can be performed in the Fourier domain. In optics, this is a classical Fourier processor with one optical filter in the Fourier plane of a 4-\textit{f} system (see Fig.~\ref{fig:teaser}C). The trainable parameters of the model are the transmission coefficients of the Fourier filter. For a given input field with an ‘$N$’ number of upstream diffraction-limited spots, the filter consists of ‘$2N$’ trainable parameters. From a machine learning standpoint, we can call this a “differentiable optical architecture” with $2N$ parameters. Note that our architecture was derived with the help of desired “inductive biases” (linearity, complex weights, and translation invariance). We also note that one could further generalize this idea to other optical processors like D2NNs by ignoring the inductive biases and having the model learn them through data at the expense of increased parameters (see Fig.~\ref{fig:teaser}D). Nevertheless, both these models are examples of potential “optical architectures” and can be used to initialize an optical design top-down. In this work we discuss how they are used in our proposed framework; we leave the discovery of better architectures for future work. 

The last piece of the puzzle is how to configure or “learn” the parameters of the optical architecture through data. If we have a dataset of paired inputs and outputs, one may start with a randomly initiated optical architecture and optimize its parameters given a loss function reflecting the desired image translation behavior of the system. To do this, we need a suitable loss function to optimize, and a dataset with paired inputs and outputs. Let’s discuss these two aspects with respect to our example. First, in our example of all-optical phase retrieval microscopy, we want to measure the phase information of the input light field on the camera of the optical system (i.e. the output). In other words, the output light field’s intensity should contain the phase information of the input field. So, the loss function should reflect this behavior. Thus, any loss that measures the distance between a scaled version of the output intensity and the input phase is suitable. In other words, the loss should optimize for the output intensity to be directly proportional to the input phase. We discuss the losses used in this work in subsection~\ref{subsec:method-losses-learning}. Second, we need a dataset with input-output pairs to train the model. Such a dataset is easy to formulate with only the input light fields. This is because the desired output is readily available in the inputs, i.e., the phase component of the input light field. In this work, we synthetically generated input light fields for objects such as MNIST digits by placing them in the phase components of the light field. We also utilized pre-acquired light fields of HeLa cells, and bacteria specimens. We discuss these datasets in section~\ref{sec:results}. 

Last, this process can be mathematically formulated as below. 
\begin{equation}
\label{eq:math_formulation}
A_{out} e^{j\phi _{out}} =G_{\{w_{i}\}}\left(A_{in} e^{j\phi _{in}}\right)\\
\end{equation}
Here, $G_{\{w_i\}}(.)$ is the optical model with a set of learnable parameters $\{w_i\}$. $A_{in}  e^{j\phi_{in}}$ is the input light field with $A_{in}$ and $\phi_{in}$ as the amplitude and phase components, respectively; similarly,  $A_{out}  e^{j\phi_{out}}$ is the output light field. We then utilize a loss function $\mathcal{L}(|A_{out} |^2/S,\phi_{in})$ that imposes the output intensity, $|A_{out}|^2$, to be directly proportional to the input phase $\phi_{in}$. Here $S$ is the constant of proportionality which is either pre-set or learned (see section~\ref{subsec:method-losses-learning}). Finally, we perform the optimization over the training dataset as

\begin{multline}
\{w_{i,\text{optimal}}\} = \underset{\{w_{i}\}}{\arg\min}\Bigg(
\quad \underset{\forall \phi_{in} \in \text{Trainingset}}{\textit{Average}} \Bigg( \\
\quad\quad \mathcal{L} \left( \frac{|G_{\{w_{i}\}} \left( A_{in} e^{j\phi_{in}} \right) |^{2}}{S}, \phi_{in} \right)
\Bigg)
\Bigg).
\end{multline}

\input{sec/big_table_3}

\subsection{All-optical Phase to Intensity Conversion as a Learning Problem}
\label{subsec:method-losses-learning}

In a machine learning perspective, all-optical phase to intensity conversion is an image translation task. A computer model, subject to the physics of light propagation (\emph{the rules}), should learn to translate complex-valued optical fields (inputs), to intensity maps proportional to the input phase (outputs). This task should be embedded in a loss function (\emph{the questions}) and training data (\emph{the correct answers}). 

To this end, we first introduce \emph{phase reconstruction loss}, $\mathcal{L}_{\phi}$. An input optical field, $x_{in}= A_{in}e^{j\phi_{in}}$ (where $A_{in}$ is the amplitude and $\phi_{in}$ is the phase of the input optical field) is propagated through the proposed model $G_{w_i}(.)$ to produce the output field $x_{out} = A_{out}e^{j\phi_{out}} = G_{w_i}(x_{in})$ (where $A_{out}$ is the amplitude and $\phi_{out}$ is the phase of the output optical field). Then phase reconstruction loss is defined as,

\begin{equation}
    \label{eq:phase_recon}
    \mathcal{L}_{\phi}=\mathbb{E}_{x_{in} \sim P_X}\left[Rh_{\delta}(|A_{out}|^2, \phi_{in}/ k)\right],
\end{equation}
where $k$, $P_X$, and $Rh_{\delta}(.)$ respectively represent the maximum phase of a phase object of the considered dataset, the probability distribution of phase objects, and the below-defined reverse Huber loss \cite{Zwald2012TheEffect}. 
\begin{equation}
Rh_{\delta } (y,f( x) )=\begin{cases}
|y-f(x),| & \text{ for } |y-f(x)|\leq \delta \\
\frac{(y-f(x))^{2} +\ \delta ^{2}}{2\delta, } & \text{ otherwise}.
\end{cases}
\end{equation}
We chose the reverse Huber loss over L1 or MSE loss as it produces better reconstruction quality (based on the empirical results). We set the $\delta$ threshold to 0.95 of the standard deviation of the ground truth image following a similar procedure to \cite{Mengu2021}. In $\mathcal{L}_{\phi}$ however, the constant of proportionality of the objective (i.e., of $|A_{out}|^2 \propto \phi_{in}$) is fixed. We propose a second \emph{learned transformation loss}, $\mathcal{L}_{LT}$, which relaxes the model to learn a constant of proportionality ($S$). 
\begin{equation}
    \label{eq:lt_loss}
    \mathcal{L}_{LT}=\mathbb{E}_{x_{in} \sim P_X}\left[Rh_{\delta}\left(\frac{|A_{out}|^2}{S}, \frac{\phi_{in}}{k}\right)\right],
\end{equation}
$\mathcal{L}_{LT}$ is more consistent with the problem formulation in the literature \cite{Gluckstad1996}. Results of learning with these losses are summarized in supplementary materials.


\subsection{Complex-valued Linear CNN: Feasibility of Linear Phase Retrieval Using a Black-box Model}
\label{subsec:feasibility-linear-PR}

All-optical phase to intensity conversion is analytically unsolvable for an arbitrary input. We, therefore, first established the existence of approximate solutions to our data distributions. To do that, we trained a complex-valued linear CNN (C-CNN) imitating optical constraints. We experimented with different hyper-parameters (the number of layers, kernel size, and the number of channels) in the CNN architecture for each dataset (presented in supplementary results). All kernels were complex-valued \cite{trabelsi2018deep} with no non-linear activations. A bias term was added only to the last layer. The CNN architecture is shown in Fig. \ref{fig:teaser}B. The results of this architecture are discussed in section~\ref{sec:results}.

\subsection{Learnable Fourier Filter: Black-box to an Optical Architecture}
\label{sec:learned_filter}


In the learnable Fourier filter (LFF) model, we optically implemented the convolution operation. Fig.~\ref{fig:teaser}C shows the optical schematic of the model. The model consisted of an optical 4-\textit{f} system and a circular Fourier filter. The filter was on a $256 \times 256$ grid of  transmission coefficients. The coefficients inside the filter were treated trainable; the ones outside were set to zero (see filters in Fig.~\ref{fig:overall_summary}F1). The input/output of the model was of the same size as the filter. But the information of the phase object was placed within the small circular aperture on a $32 \times 32$ grid. The grid was padded $8 \times$ to make up the $256 \times 256$ input to maintain a sufficient frequency sampling according to the Nyquist criterion. The first lens of the 4-\textit{f} system, Fourier-transforms the input field; the filter modulates the field; and the second lens, inverse-Fourier-transforms the field back to the spatial domain. This physical system can be mathematically modeled as, 
\begin{equation}
\label{eq:fourier_model}
G_{\{H_t(x,y)\}} \equiv IDFT2(H_t(x,y) \circ DFT2(\cdot) )
\end{equation}
where $H_t$ represents the learnable transmission coefficient matrix.  $DFT2(\cdot)$ and $IDFT2(\cdot)$ are the two-dimensional Fourier and inverse Fourier transforms. The coefficients of the filter were initialized and optimized on each training data set, through the loss functions in section~\ref{subsec:method-losses-learning}. We conducted multiple experiments with different configurations and parameterizations of learnable Fourier filters on \emph{each} dataset as explained in section~\ref{sec:results}.

\subsection{D2NN: All-optical Phase Retrieval Through an Optical Architecture without Inductive Biases}
\label{sec:phased2nn}

Our third model is a D2NN. The model consists of a large number of parameters distributed in a partially-connected cascade of diffractive layers. A stand-alone layer is functionally similar to the learnable filter presented in the previous subsection. But their placement and combined function are not driven by any inductive biases. Thus the model should learn these constraints through data. Nevertheless, the D2NN is an interesting miniaturized design.

The proposed architecture is shown in Fig.~\ref{fig:teaser}D. The network consisted of 8 diffractive layers which were separated by $5.3\lambda$ distance ($\lambda$ is the operating wavelength of the input field). Each layer of the network consisted of a $128 \times 128$ neuron grid that contained trainable complex-valued transmission coefficients. The size of each neuron was $\lambda/2 \times \lambda/2$. The input to the network is a light field, i.e., a complex-valued image, where the information of interest can be found in its phase. This input was located in a $32 \times 32$ grid, $5.3\lambda$ before the first layer of the D2NN. The reconstruction plane (i.e., the image plane where a detector is placed) was at a $9.3\lambda$ distance from the last optical layer. The reconstruction was done on $32 \times 32$ grid on this reconstruction plane. The element size of the input grid and reconstruction grid was the same as the size of a neuron. Hence, this compact system has dimensions of $64\lambda \times 64\lambda \times 51.7\lambda$. 

After mathematically modeling the diffractive layers as described in subsection~\ref{sec:prelim_d2nn}, the learnable transmission coefficients ($t_i$) of each layer were optimized with the objective defined in subsection~\ref{subsec:method-losses-learning}. To maintain the passive nature of the layers, we constrained the amplitudes of the transmission coefficients to the range $[0,1]$. We note that, with no inductive biases and a large number of learnable parameters, D2NN is an optical architecture at one extreme end of the spectrum.


%% file: sec/big_table_3.tex
\begin{table*}[t]

\centering
\caption{Overall quantitative results. Performance of the Complex-valued CNN (C-CNN), GPC method, the learnable Fourier filter (LFF), and the D2NN are shown for each dataset. For the models learned with the $\mathcal{L}_{LT}$ loss (see section~\ref{subsec:method-losses-learning}), the best-performing cases in each performance metric for each dataset are in bold. Same values are compared side-by-side in Fig. \ref{fig:overall_summary}-H1.}
\label{tab:overall_results}
\resizebox{\textwidth}{!}{
\begin{tabular}{lccccccccccccccc}
Method & \multicolumn{3}{c}{MNIST $[0, \pi]$}  & \multicolumn{3}{c}{MNIST $[0, 2\pi]$} & \multicolumn{3}{c}{HeLa $[0, \pi]$} & \multicolumn{3}{c}{HeLa $[0, 2\pi]$} & \multicolumn{3}{c}{Bacteria $[0, \pi]$} \\ \cline{2-16} 
 & SSIM $\uparrow$ & L1 $\downarrow$ & {\color{black}PSNR} $\uparrow$ & SSIM $\uparrow$ & L1 $\downarrow$ & {\color{black}PSNR} $\uparrow$ & SSIM $\uparrow$ & L1 $\downarrow$ & {\color{black}PSNR} $\uparrow$ & SSIM $\uparrow$ & L1 $\downarrow$ & {\color{black}PSNR} $\uparrow$  & SSIM $\uparrow$ & L1 $\downarrow$ & {\color{black}PSNR} $\uparrow$ \\ \midrule[1.5pt]   

C-CNN & 0.9982 & 0.0041 & {\color{black}40.16} & 0.7913 & 0.0539 & {\color{black}17.28} & 0.9417 & 0.0248 & {\color{black}29.76} & 0.8619 & 0.0485 & {\color{black}20.67} & 0.9938 & 0.0008 & {\color{black}51.22} \\ \cmidrule[0.5pt]{1-16}

\rowcolor{gray!10}
LFF & 0.9205 & \textbf{0.0196} & {\color{black}\textbf{27.90}} & 0.6814 & 0.0608 & {\color{black}17.02} & 0.7783 & \textbf{0.0688} & {\color{black}\textbf{20.05}} & \textbf{0.6078} & \textbf{0.0937} & {\color{black}\textbf{16.86}} & 0.9823 & 0.0010 & {\color{black}49.40} \\

GPC$^{\star}$ & 0.9036 & 0.0320 & {\color{black}23.35} & 0.4406 & 0.0974 & {\color{black}11.66} & \textbf{0.7786} & 0.0748 & {\color{black}19.21} & 0.5509 & 0.1091 & {\color{black}15.12} & 0.9600 & 0.0029 & {\color{black}42.20} \\

\rowcolor{gray!10}
D2NN & \textbf{0.9433} & 0.0201 & {\color{black}26.37} & \textbf{0.7703} & \textbf{0.0486} & {\color{black}\textbf{18.98}} & 0.6655 & 0.0936 & {\color{black}18.26} & 0.4942 & 0.1274 & {\color{black}15.05} & \textbf{0.9926} & \textbf{0.0007} & {\color{black}\textbf{52.49}} \\

\cmidrule[0.5pt]{1-16}

$\phi$-LFF & \textbf{0.9177} & \textbf{0.0196} & {\color{black}\textbf{27.90}} & 0.6777 & 0.0608 & {\color{black}17.05} & \textbf{0.7771} & \textbf{0.0688} & {\color{black}\textbf{20.06}} & \textbf{0.6096} & \textbf{0.0937} & {\color{black}\textbf{16.86}} & 0.9825 & 0.0010 & {\color{black}49.31} \\

\rowcolor{gray!10}
$\phi$-LRF & 0.8573 & 0.0293 & {\color{black}25.16} & 0.6211 & 0.0671 & {\color{black}16.60} & 0.7583 & 0.0730 & {\color{black}19.48} & 0.6078 & 0.0943 & {\color{black}16.77} & 0.9508 & 0.0023 & {\color{black}45.18} \\

$\phi$-GPC$^{\star}$ & 0.8466 & 0.0430 & {\color{black}21.32} & 0.4191 & 0.0982 & {\color{black}11.67} & 0.7297 & 0.0755 & {\color{black}19.48} & 0.5184 & 0.1100 & {\color{black}15.18} & 0.9479 & 0.0036 & {\color{black}41.84} \\

\rowcolor{gray!10}
$\phi$-D2NN & 0.8796 & 0.0285 & {\color{black}24.17} & \textbf{0.6906} & \textbf{0.0591} & {\color{black}\textbf{18.00}} & 0.6334 & 0.0945 & {\color{black}18.16} & 0.4191 & 0.1316 & {\color{black}14.79} & \textbf{0.9867} & \textbf{0.0009} & {\color{black}\textbf{49.72}} \\   

 \midrule[1.5pt]

\multicolumn{10}{l}{$^{\star}$ \footnotesize Output is scaled for each dataset (see supplementary section B.1)}

\end{tabular}}
\end{table*}

%% file: sec/5_results.tex
\section{Experiments Results and Discussion}
\label{sec:results}
We present experimental results in this section. to this end, we present the datasets employed in the experiments in the next subsection.
\subsection{Datasets}

\paragraph{Phase MNIST Digits}
To preserve consistency with existing studies~\cite{Lin2018} we consider two \textit{Phase MNIST digit} datasets for evaluation. To construct these datasets, we convert the images from the MNIST digit dataset to phase images ($e^{j\phi _{in}}$) such that $\phi _{in} \in [ 0,\ \pi ]$ for one dataset (MNIST$[0, \pi]$), and $\phi _{in} \in [ 0,\ 2\pi ]$ for the other dataset (MNIST$[0, 2\pi]$). Each dataset contains 54000, 6000, 10000 train, validation, test images, respectively. We considered these datasets as \emph{sparse datasets} because they contain relatively less complex structures (\eg, contains only a single digit located at the center of the image grid).
 
\paragraph{HeLa Dataset}
To evaluate the proposed method on microscopy, we acquire a dataset of \textit{HeLa} cells where the samples are illuminated with 600 nm light. Dataset contains 10344 images where each image represents a complex field. 80\%, 10\%, 10\% of the dataset are utilized for training, validation, test sets, respectively. The electric field amplitudes are normalized to [0, 1] based on the maximum amplitude obtained through the steps described in the quantitative phase imaging section in the supplementary materials. To remove outliers in the phase, values are clipped to $[0, 2\pi]$. To study the effect of the distribution of phase values in a limited range, the phase information of the above dataset is re-scaled to $[0, \pi]$ and considered as a separate dataset for reporting performance. More details on the sample preparation are presented in supplementary materials. 

\paragraph{Bacteria Dataset}
This is another acquired dataset for the purpose of evaluating the network performances for microscopy. The dataset has phase information $\phi _{in} \in [ 0,\pi ]$. This can also be considered as a \textit{sparse dataset} similar to the Phase MNIST dataset. More details on the sample preparation are presented in supplementary materials. 

\begin{figure*}[htbp]
    \centering
    \includegraphics[width=.9\textwidth]{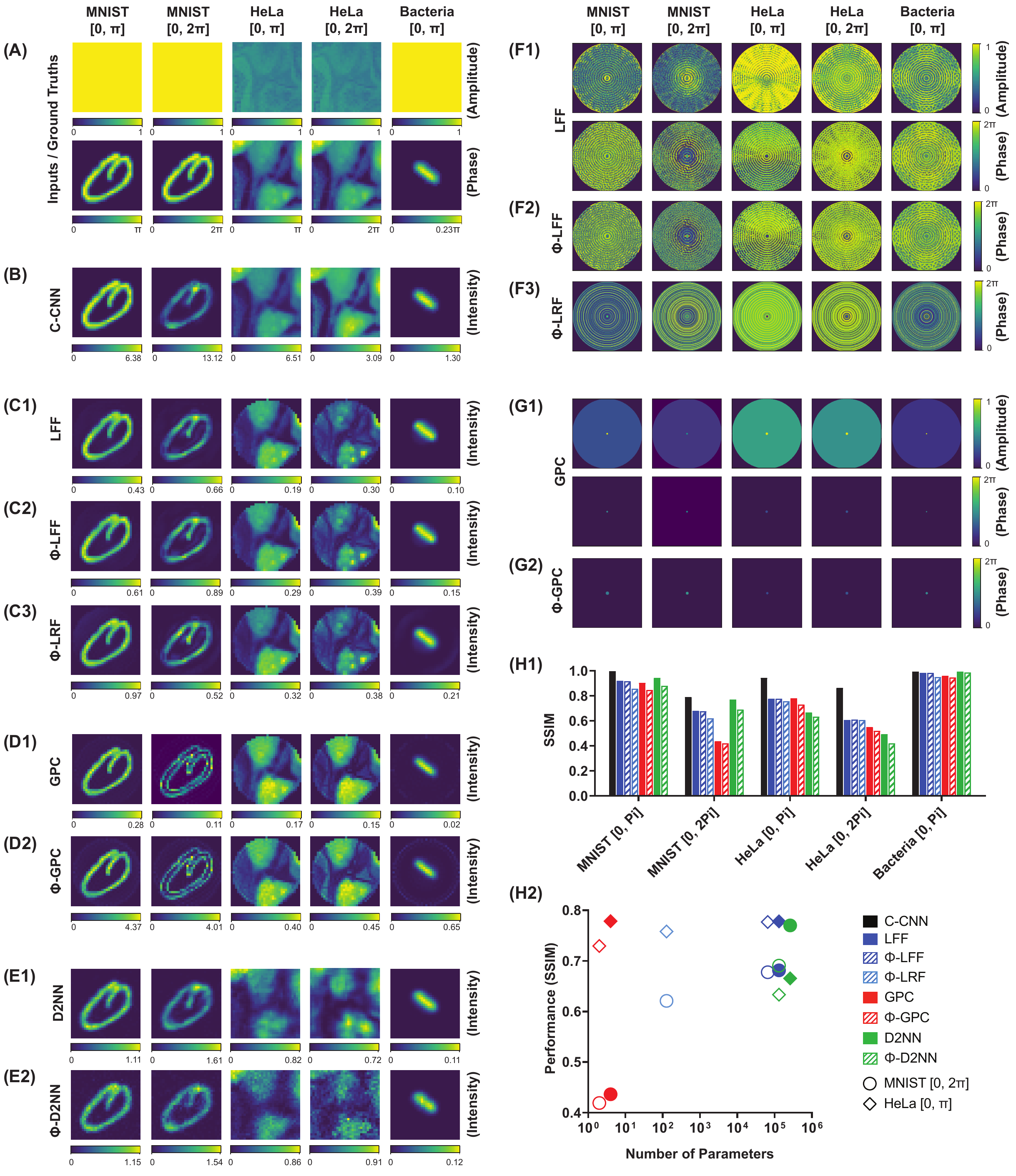}
    \caption[short]{\textbf{Overall qualitative results.} \textbf{(A)} The amplitude (top row) and phase (bottom row) of the input field to the system for each dataset. \textbf{B} The results of the complex-valued linear CNN for the phase-to-intensity conversion task on each test dataset. The output intensity maps are visually similar to the input phase maps, suggesting a linear model can learn to convert phase to intensity for a given dataset. \textbf{C1, C2, C3} Show the reconstruction results by the best performing LFF, $\phi$-LFF, and $\phi$-LRF learned for each dataset, respectively. Similarly, \textbf{D1, D2, E1, E2} show the respective results for the GPC, $\phi$-GPC, D2NN, and $\phi$-D2NN. \textbf{F1} Shows the amplitude (top row) and phase (bottom row) of the learned LFFs for each dataset. \textbf{F2, F3}  Shows the phase of the learned $\phi$-LFF and $\phi$-LRF for each dataset respectively. \textbf{G1} Shows the amplitude (top row) and phase (bottom row) of the learned GPC filters for each dataset. \textbf{G2} Shows the phase of the learned $\phi$-GPC filter for each dataset. \textbf{H1} Shows a comparison of $\partial\mu$ architectures on different datasets. Our approach consistently outperforms the GPC~\cite{Gluckstad2009} for all datasets except the HeLa$[0,\pi]$ dataset, where it performs on-par with the GPC. The complex-valued CNN sets the empirical upper bounds for each dataset. \textbf{H2} Shows the performance of each $\partial\mu$ and GPC architecture with the trainable number of parameters. The $\phi$-LRF achieves comparable/ superior performance for the MNIST $[0,2\pi]$ and HeLa $[0, \pi]$ datasets with a significantly lower number of trainable parameters compared to the LFFs and $\phi$-D2NN architectures. Even though the GPC shows comparable performance with the $\phi$-LRF for the HeLa $[0, \pi]$ dataset, the $\phi$-LRF shows a significant improvement in performance for the MNIST $[0, 2\pi]$ dataset.}
    \label{fig:overall_summary}
\end{figure*}

\subsection{Implementation Details}
We implemented the optical networks with Python version 3.6.13. We considered auto differentiation in PyTorch\cite{pytorch} framework version 1.10.0 for the training of the optical networks. We conducted all experiments on a server with 12 Intel(R) Xeon(R) Platinum 8358 (2.60 GHz) CPU Cores, an NVIDIA A100 Graphical Processing Unit with 40 GB memory running on the Rocky 8 Linux distribution.
We considered a training batch size of 32 examples and a learning rate of 0.01 for the networks. We used Adam\cite{adam} as the optimizer for all optimizations.

\subsection{Evaluation Metrics}

For all of our experiments we utilized the L1 loss, {\color{black} peak signal-to-noise ratio (PSNR)} and structural similarity index (SSIM)\cite{ssim_paper} to evaluate our output intensity reconstruction compared to the input phase information. L1 loss measures the pixel-wise difference in phase values, and can be treated as a metric for quantitativeness. {\color{black} PSNR quantifies the ratio between the maximum possible power of the signal and the power of corrupting noise affecting the reconstruction.} SSIM is a commonly used metric to measure the perceived image quality among the image processing and computer vision communities.

{\color{black}

\subsection{Assumptions and Scope Definition}
\label{sec:discussion_assumptions_scope}

While $\partial\mu$ is a top-down optical design approach, the capabilities of the learned microscope are strictly governed by the underlying physics of the forward model and the information content of the training data.

\paragraph{Assumptions and In-Scope Tasks} The current implementation assumes a linear, shift-invariant optical system, which is valid for the demonstrated task of phase retrieval in thin, weakly scattering samples limited by the numerical aperture. 
\paragraph{Out-of-Scope Problems} The current linear formulation cannot address strongly non-linear regimes, such as multiple scattering in thick tissue (descattering), nor can it recover full 3D information from a single 2D acquisition without extending the forward model and acquisition strategy (e.g., tomography). Furthermore, it cannot achieve resolution beyond the diffraction limit imposed by the numerical aperture unless specific priors or non-linear light-matter interactions are incorporated into the design. We leave investigation of such demonstrations for future work.
}

\subsection{Evaluating Feasibility of Linear Phase Retrieval and Setting Empirical Upper Bounds}

Qualitative results of the CNN model are shown in Fig.~\ref{fig:overall_summary}B. Corresponding quantitative results are shown in row C-CNN in  Table~\ref{tab:overall_results}. As seen, the model achieved high SSIM\cite{ssim_paper} values for all datasets. For MNIST$[0, 2\pi]$ and HeLa$[0, 2\pi]$ datasets the SSIM values were slightly lower than those for the others. This is consistent with previous work (chapter 2 of~\cite{Gluckstad2009}) that implies that the linear approximation does not hold well at large phase variations. Qualitative results agree with this observation (see Fig.~\ref{fig:overall_summary}B). Thus our complex-valued CNN, established the existence of linear approximate solutions for given data distributions, while also setting empirical upper bounds for all-optical phase retrieval on each dataset. 

\subsection{Experiments with Optical Top-down Networks}

\paragraph{Learnable Fourier Filter}
For each dataset, we tested several configurations and parameterizations of learnable Fourier filters. First, all transmission coefficients (both amplitude and phase) of the LFF were treated trainable. Results for this configuration are shown in row LFF in Table~\ref{tab:overall_results} and Fig.~\ref{fig:overall_summary}.C1. The filters learned, for each dataset, are shown in Fig.~\ref{fig:overall_summary}.F1.  Second, only the phase components of the transmission coefficients were treated trainable; the amplitudes were frozen to a unit amplitude. The results for this configuration are shown in row $\phi$-LFF in Table~\ref{tab:overall_results} and Fig.~\ref{fig:overall_summary}.C2. The filters learned, for each dataset, are shown in Fig.~\ref{fig:overall_summary}.F2. Third, we constrained the $\phi$-LFF to be radial symmetric, to promote rotational invariance as an additional inductive bias. This configuration allowed us to reparameterize the filter design with  two to three orders of magnitude lesser number of parameters compared to the $\phi$-LFFs. Results for this configuration are shown in row $\phi$-LRF in Table~\ref{tab:overall_results} and Fig.~\ref{fig:overall_summary}.C3. The filters learned, for each dataset, are shown in Fig.~\ref{fig:overall_summary}.F3. As seen in our results, LFF performance decreased from the C-CNN baseline for all datasets and the degradation was more pronounced for the two Hela datasets. Interestingly $\phi$-LFF performed similarly to the LFF models. Last, $\phi$-LRF models performed slightly worse than the $\phi$-LFF and LFF models. {\color{black} Furthermore, we analyzed the robustness of the trained LFF models against system non-idealities such as phase quantization and parameter perturbations; detailed results are presented in Supplementary Material Section E.}

\paragraph{D2NN}
Next, we tested an optical architecture that does not use any inductive bias for the same all-optical phase retrieval task. In our study, we consider two types of D2NN designs; D2NNs composed of layers that modulate both the phase and the amplitude of the incoming waves (simply referred to as D2NN) and those that modulate only the phase (referred to as $\phi$-D2NN). Quantitative results of D2NN and $\phi$-D2NN designs are shown in Table~\ref{tab:overall_results}. Qualitative results are shown in Fig.~\ref{fig:overall_summary}.E1, and Fig.~\ref{fig:overall_summary}.E2. As seen in the results, D2NNs performed slightly better than LFFs for MNIST and bacteria datasets but worse than LFFs for the more complex Hela datasets. Unlike in LFFs, $\phi$-D2NNs performed slightly worse than D2NNs.

None of the above architectures could reach the numerical upper bound (i.e., the C-CNN performances),  perhaps due to sub-optimal convergence of parameters. Thus an interesting future direction is to explore better optimizers for these physics-constrained optical architectures.  Nevertherless, this sub-par performance does not necessarily mean that the results are not quantitative since L1 loss can be treated as a metric for quantitativeness. However, in the all-optical setting, there is a power depended constant of proportionality that should be measured to convert the intensity values to the exact phase values. Therefore another interesting future direction is to develop a robust strategy to measure the propotionality constant, perhaps through a second intensity measurement.

Moreover, it is interesting that all models performed similarly on datasets with simple features and low phase variations (i.e., the MNIST$[0,\pi]$ and bacteria$[0,\pi]$). For Hela$[0,\pi]$, with complex features but low phase variation, D2NN designs performed much worse than others most likely due to convergence issues. For  Hela$[0,2\pi]$, with complex features and high phase variation, LFFs performed considerably better than the rest. In general, we conclude that LFFs struck a balance between the number of parameters and performance in most cases (see Fig. \ref{fig:overall_summary}.H2). This result suggests that a single diffractive layer, when placed right, can outperform a D2NN. This is because, though comprised of many layers, a D2NN is a cascade of linear operators that collapses to a single linear operation. The cascading is needed in the spatial domain to connect all pixels; but the same can be achieved by the Fourier operator. In fact, the Fourier operator fully connects the network while D2NN is partially connected. Furthermore, compared to D2NNs, it is easier to train LFFs due to the reduction of trainable parameters and the reduced computational complexity in wave propagation. For instance, while D2NNs are supposed to learn the inductive bias through data, they are poor in doing so in the presence of complex features. Moreover, our Phase LRF parameterization, in comparison to the other LFFs and D2NNs (Fig. \ref{fig:overall_summary}.H2), can reduce the number of parameters by orders of magnitudes, at the cost of slight performance reduction. This may be a promising path toward efficient training.

We would also like to note that the top-down designed models will not generalize beyond their data distribution.  It is important to understand the limits of the model performance when operated beyond the intended data distribution. To explore the generalizability limits of the LFF and D2NN, we conducted a series of validation experiments and presented in supplementary materials. Specifically, we trained the proposed $\phi$-LFF, $\phi$-LRF, and $\phi$-D2NN on a specific dataset and tested them on the rest of the datasets. We observe that the models trained on the datasets with large phase variation tend to generalize on the same datasets with smaller phase variation.

\subsection{Generalized Phase Contrast: All-optical Phase Retrieval by an Analytically Designed Optical Architecture}
\label{sec:GPC}
In this subsection, we tested a fully analytically designed optical architecture with only a few learnable parameters. Optical architectures presented in the previous sections are top-down approaches that use data to learn the optical design. In these models, a large number of parameters should be learned in general. In this subsection, we compare the other end of the spectrum, where the entire design is done bottom-up with domain knowledge. To this end, we tested an all-optical phase retrieval method called GPC~\cite{Gluckstad1996}. In GPC a two-part Fourier filter is designed to generate interference with the near-zero frequency components of the object light field, with its high-frequency components. The passband and transmission coefficients of the two parts of the filter were designed using the target data distributions. This filter is characterized by four parameters ($A,B,\theta,\Delta f_r$), and we followed two methods to select the best GPC filter parameters for a given input data distribution. The detailed guidelines of the design process are explained in supplementary materials.

Similar to the LFF case, we tested both GPC and $\phi$-GPC (phase-only filter) configurations. Results are shown in Table~\ref{tab:overall_results} and Fig.~\ref{fig:overall_summary}.D1 \& D2. The filters learned, for each dataset, are shown in Figs.~\ref{fig:overall_summary}.G1 and G2. GPC and $\phi$-GPC performed similarly to LFF and $\phi$-LFF models on MNIST $[0, \pi]$, Hela $[0, \pi]$, and Bacteria $[0, \pi]$ datasets. But GPC models were considerably worse than LFF models for MNIST $[0, 2\pi]$ and Hela $[0, 2\pi]$ datasets. In conclusion, at-least one top-down design (i.e., from LFF, D2NN, $\phi$-LFF, $\phi$-LRF, $\phi$-D2NN) \emph{outperformed} their bottom-up GPC and $\phi$-GPC counterparts  in terms of L1-loss on all datasets. SSIM trended the same except on HeLa $[0, \pi]$. This is because top-down design approaches could utilize the data distribution apriori while bottom-up design {\color{black}approaches} cannot. But this advantage comes at the cost of having to effectively optimize a large number of parameters.

Moreover, we observed that some LFF designs contained similar features to the GPC design. Specifically, most LFF, $\phi$-LFF, and $\phi$-LRF designs learned to delay near-zero frequencies with a distinct phase shift in the middle. This is not surprising given we are aware of the rules used to design GPCs. Nevertheless, it may be of interest to optical designers in terms of discovering new design rules that aren't immediately apparent. In a hypothetical world where GPC had not been discovered, one may be able to analyze our learned LFF designs and simplify them through logical ablation experiments until an analytical rule can be distilled out of the learned designs. While we do not claim $\partial\mu$ can discover new interpretable design rules, we do not reject the possibility of a refined $\partial\mu$ framework that can.

\subsection{Experimental Validation of the Learnable Fourier Filter}
\label{sec:lff_experiment}
We now present an experimental validation of the LFFs to show that our top-down designed optical models can be experimentally implemented. The goal was to validate that our data-driven top-down designs can be physically built with intended properties. To this end, we implemented the $\phi$-LFF design for the MNIST dataset using a spatial light modulator (SLM). We fabricated a phase mask of digit 7 (from the test set) as the test object (using implosion fabrication ~\cite{Oran2018}). The optical schematic of the setup is shown in Fig.~\ref{fig:lff_experiment}.A (a photograph of the experimental setup is also shown in the supplementary material). An optically pumped semiconductor laser (Sapphire LPX 488, Coherent, Inc, CA, USA),  was used as the light source. A half-wave plate ensured the correct polarization for the SLM. The output beam was expanded and collimated using a beam expansion system. A reflective neutral density (ND) filter (ND 2.0, Thorlabs, USA) was placed in the beam path to control the laser intensity. The phase object (mounted on a 3D stage) was illuminated using the collimated beam and the transmitted light was imaged by an objective (M Plan-50$\times$/0.75 NA, Olympus) and a tube lens ($f=180$ mm, Achromatic, Thorlabs, USA). An aperture was placed at the image plane to match the field of view to that of the design. Next, a reflective 4-\textit{f} system was implemented using a single lens ($f=180$ mm, Achromatic, Thorlabs, USA) and the SLM. The SLM (EXULUS-HD2, Thorlabs, USA) was placed at the Fourier plane of the 4-\textit{f} (see Fig.~\ref{fig:lff_experiment}.A). The reflected beam passing through the same lens was projected on the detector (Grasshopper 3, FLIR, Canada) using a beam splitter (BS 50/50, Thorlabs, USA). The respective LFF (see Fig.~\ref{fig:lff_experiment}.C) was configured on the SLM after resampling to match the design (see supplementary materials for details).  Fig.~\ref{fig:lff_experiment}.D shows the image recorded on the camera, i.e., the output intensity.  Fig.~\ref{fig:lff_experiment}.B2 shows the expected phase image (from the fabricated digit) at the input image plane.   

This demonstration/experimental validation was strictly qualitative and intended to be only a proof of concept. Future work should  experimentally replicate the numerical benchmarks presented in this work over multiple datasets. In our proof of concept experiment, we observed a model mismatch. Our model did not include exact discretization and quantization constraints imposed by the SLM and the camera detector at the design stage. We introduced these constraints post-design, which lead to the model mismatch (discussed further in the supplementary materials).  A potential future direction is to build discretization- and quantization-aware differentiable models. One may also explore fabricating the LFF designs as phase and intensity masks. Nevertheless, the results presented here demonstrate that our top-down designs do in fact work as intended in experiments.


\begin{figure}[t]
\begin{center}
\includegraphics[width=1.0\columnwidth]{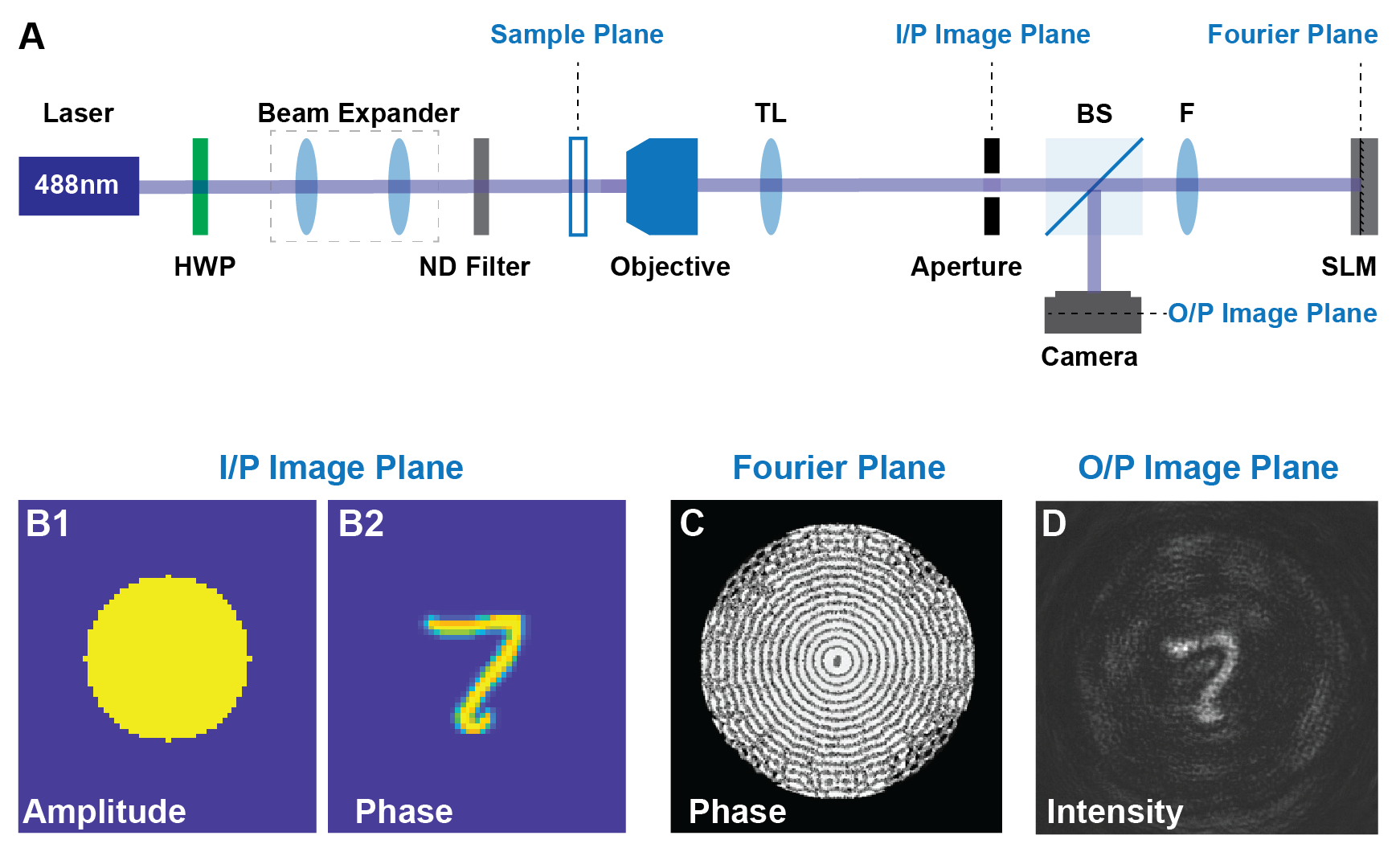}
\end{center}
\caption{
\textbf{Experimental validation of the LFF using an SLM.} \textbf{(A)} The optical schematic of the experimental setup (HWP - half-wave plate; ND Filter - neutral density filter; TL - tube lens; BS – 50/50 beam splitter; F – 180 mm focal length lens; SLM spatial light modulator; I/P input; O/P – output). \textbf{(B1,2)} The ideal input image field of the fabricated phase object (B1 – Amplitude, and B2 – phase) generated at the first image plane of the microscope. \textbf{(C)} The phase filter configured on the SLM. \textbf{(D)} The image recorded by the camera (i.e., the intensity) at the output image plane of the 4-\textit{f} system.}
\label{fig:lff_experiment}
\end{figure}

%% file: sec/6_conclusions.tex


\section{Conclusion}
\label{sec:con}
The current {\color{black}workflow} within microscopy is to develop the system hardware for a specific microscope (e.g., QPM) and then use that for different applications. Here, we propose to {\color{black} augment this process} by introducing the concept of differentiable microscopy ($\partial\mu$) that can learn the optical imaging process required for a given technical specification (e.g., extracting phase information) and for different targeted applications. As a first demonstration of $\partial\mu$ we learned all-optical designs for phase retrieval microscopy.

In conventional QPM, the imaging pipeline is all-optical, but the imaging reconstruction requires computation resources. Our learned phase retrieval microscopes completely remove the need of post-imaging computation resources and have a huge potential of making the footprint of the microscope compact.  We believe this work will motivate researchers to exploit our designs for applications where high-speed, high-throughput, and compact setups are needed. For instance, our designs may find use in point-of-care systems in micro-biology \cite{Jo2014}, histopathology \cite{azeem2021}, and  material sciences \cite{Zhou2015}. Moreover, the proposed all-optical phase-to-intensity conversion can be exploited for applications beyond microscopy,  such as for two-photon-polymerization based nano-fabrication~\cite{Gluckstad2009} and optogenetics~\cite{optogenetics}.  {\color{black}We believe this approach provides a practical methodology for integrating data-driven optimization into the optical design process.}
Notably, some of our learned designs are similar to the GPC concept \cite{Gluckstad2009}.  It would be interesting to see if $\partial\mu$ can invent well-known microscope designs from data for various other tasks (such as for depth-resolved imaging or for optical coherence tomography). We anticipate that differentiable microscopy will open doors to a new generation of interdisciplinary scientists, to design optical systems for challenging imaging problems.




%% file: sec/X_supplementary_v2.tex

\renewcommand\thefigure{\arabic{figure}}
\setcounter{figure}{0}
\setcounter{equation}{0}

\renewcommand{\thefigure}{S\arabic{figure}}
\renewcommand{\thetable}{S\arabic{table}}
{
\centering
\Large
\vspace{1.0em}\textbf{Supplementary Material} \\
} 

\section{Preparation of HeLa and Bacteria Datasets}

\subsection{Sample Preparation:}

\paragraph{HeLa cells} HeLa cells were grown in a standard humidified incubator at 37 $^{\circ}$C with 5\% CO2 in minimum essential medium supplemented with 1\% penicillin/streptomycin and 10\% fetal bovine serum. Cells were seeded into the Polydimethylsiloxane chamber of 10 mm $\times$ 10 mm size with 150 $\mu$m thickness on a reflecting silicon substrate. The cells were left in the incubator for 1-2 days for their growth in a densely packed manner and fixed for {\raise.17ex\hbox{$\scriptstyle\sim$}}  20 min using 4\% paraformaldehyde in phosphate buffered saline. The sample is then sealed with \# 1.5 cover glass from the top which enabled to use water immersion objective lens for imaging and also avoided any air current in the specimen.

\paragraph{Bacteria} Bacteria samples has Brownian motion due to their small sizes, which introduce challenges in imaging them using a multi-shot QPM system. Multi-shot QPM acquires multiple interferometric frames for the phase recovery and therefore specimen should be immobile within the multi frame acquisition time.   
The bacteria samples were also prepared on a reflecting substrate (Si-wafer) due to the reflection geometry of our optical setup. First, the substrate was thoroughly washed with dist. H2O and dried with N2 gas. The 2-polydimethylsiloxane (PDMS) chamber of opening $10 \times 10$ mm was placed on top of Si-wafer and the opening area is filled with poly-L-lysine (PLL). The substrate is incubated with PLL for 15-20 minutes and then access amount of PLL is removed and gently washed using phosphate-buffered saline (PBS). The thin layer of PLL positively charges the Si-wafer surface and helps to adhere the negatively charge bacteria cells for imaging. The 20 - 25 $\mu$l volume of bacteria sample is pipetted in the PDMS chamber and left for 30 minutes of incubation. The bacteria cells were adhered to the surface of Si-wafer due to the electrostatic attraction between the negatively charged bacteria and positively charged wafer surface. The immobile bacteria cells were further gently washed off with PBS and covered with $\#$ 1.5 cover glass from the top to use high resolution water immersion objective lens (60$\times$/1.2NA) for imaging.

\subsection{Quantitative Phase Imaging:}
\label{sec:qpi_hela}
For quantitative phase imaging, HeLa sample is placed under the Linnik interferometer based quantitative phase microscopy (QPM), which works on the reflection mode. The sample is illuminated with partially spatial and highly temporal coherent light source also called pseudo-thermal light source to generate high quality interferometric images. This light source illumination has unique advantages such as high spatial and temporal resolution, high spatial phase sensitivity, extended range of optical path difference adjustment between the interferometric arms etc over conventional light sources, e.g., lasers, white light, and light emitting diodes \cite{azeem2021}. More details of the experimental setup can be found in previous work \cite{azeem2021}.
The imaging of HeLa cells is performed using high resolution water immersion objective lens (60$\times$/1.2NA) at 660 nm wavelength. This provided the theoretical transverse resolution approximately equal to 275 nm, which is quite good for high resolution phase recovery of the specimens. The interferometric data is recorded with 5.3 Megapixels ORCA-Fusion digital CMOS camera (model \# C14440-20UP). The camera sensor has 2304 $\times$ 2304 pixels with 6.5 $\mu$m pixel size and provided fairly big 225 $\mu$m $\times$ 225 $\mu$m FOV at 60$\times$ optical magnification in QPM system. We acquired approximately 500 such FOVs of different region of interests of the HeLa sample and each FOV contained roughly 50 – 60 cells. For bacteria samples, more than 50 FOVs of different regions of interests were found to be sufficient to generate large amount of interferometric data due to their small sizes compared to HeLa cells. Each FOVs contained more than 500 bacteria cells and thus provided approximately 25000 bacteria cells interferometric images for machine learning. The recorded interferometric images are further post processed with random phase-shifting algorithm based on principal component analysis \cite{Vargas:11} for the recovery of the complex fields related to HeLa cells.

\section{Preliminaries - Mathematical Modeling of Optical Systems}

\subsection{GPC Filter Search}
\label{subsec:GPC_region_select}

We followed two methods to select the best GPC filter parameters for a given input data distribution.
\begin{itemize}
    \item \textbf{Method 1}: Obtained the best set of parameters through a grid search such that it achieves the maximum SSIM for a single batch of randomly selected ten images from the training set of a dataset.
    \item \textbf{Method 2}: Perform gradient descent using the Adam optimizer on the entire training set of a dataset to find the filter parameters.
\end{itemize}
When incorporating an SSIM loss, since the GPC method does not naturally reconstruct the output intensity ($out = |IDFT2$ $\{DFT2\{E_{in}\}\circ G(A,B,\theta,\Delta f_r)\}|$ for an input field $E_{in}$)  within the $[0,1]$ range, it is required to select the appropriate scaling factor $S$ for the filter reconstructions and incorporate it in the loss function.
Therefore, we search the suitable $S$ for the GPC reconstructions while searching for the optimal parameters that define the GPC filter, so that it can be scaled to obtain the best SSIM value as shown in eq.~\eqref{eq:ssim_best} for an intensity reconstruction within the $[0,1]$ range. From the two methods, we select and report the GPC filter that obtains the highest SSIM on the test set of a dataset.

\begin{equation}
\mathcal{L}_{SSIM} \ =\  \argmin_{A,B,\theta,\Delta f_r,S} \hspace{0.1em} -SSIM \left( \left(S*out \right) ,gt_{a}{}_{n}{}_{g}{}_{l}{}_{e} \right)
\label{eq:ssim_best}
\end{equation}

Given the size of a pixel $dx = \lambda/2$ in the spatial domain and the size of the input image $N = 256$, the frequency resolution is defined as $df = 1/(N*dx) = 1/(128\lambda)$. Table~\ref{tab:gpc_configs} demonstrates the best performing GPC and $\phi$-GPC filter configurations for each dataset.

\begin{table}[h]
\centering
\caption{\textbf{Best performing GPC and $\phi$-GPC filter configurations}}
\begin{tabular}{lccccc}
Dataset & $A$ & $B$ & $\theta$ & $\Delta f_{r}$ & $S$\\

\midrule[1.5pt]
\multicolumn{6}{l}{\hspace{4cm} {GPC Filter}}\\
MNIST $[0, \pi]$ & 0.2640 & 0.9652 & 2.8833 & 3$df$ & 3.7200 \\
MNIST $[0, 2\pi]$ & 0.1640 & 0.5360 & 3.2093 & 3$df$ & 8.2000 \\
HeLa $[0, \pi]$ & 0.5548 & 0.9747 & 1.5568 & 4.323$df$ & 3.3567 \\
HeLa $[0, 2\pi]$ & 0.5001 & 0.9843 & 1.8024 & 4.323$df$ & 3.6208 \\
Bacteria $[0, \pi]$ & 0.1600 & 0.9600 & 3.0229 & 2$df$ & 8.4400 \\

\midrule[1.5pt]
\multicolumn{6}{l}{\hspace{3.8cm} {$\phi$-GPC Filter}}\\
MNIST $[0, \pi]$ & 1.0000 & 1.0000 & 2.6852 & 6$df$ & 0.2350 \\
MNIST $[0, 2\pi]$ & 1.0000 & 1.0000 & 3.2507 & 5$df$ & 0.2050 \\
HeLa $[0, \pi]$ & 1.0000 & 1.0000 & 1.5475 & 4.166$df$ & 1.6629 \\
HeLa $[0, 2\pi]$ & 1.0000 & 1.0000 & 1.8291 & 5$df$ & 1.0000 \\
Bacteria $[0, \pi]$ & 1.0000 & 1.0000 & 2.9502 & 4.166$df$ & 0.1934 \\

\end{tabular}

\label{tab:gpc_configs}
\end{table}

\subsection{Angular Spectrum Method}

A plane wave can be considered as a combination of a set of plane waves travelling in different directions. The complex amplitudes of these plane wave components form the angular spectrum of the given wave~\cite{Ratcliffe1956}. Consider the field of a light wave propagating in the $z$ direction is given by,
\begin{equation}
    U(x,y,z,t) = U(x,y,z) e^{-j2\pi ft},
\end{equation}
where $f$ is the temporal frequency of the wave. Suppose that, at $z=0$ plane, $U(x,y,z)|_{z=0} \equiv U(0)$. It can be shown that the field $U(0)$ and its angular spectrum $A(f_x,f_y,0) \equiv A(0)$ are related by~\cite{Ratcliffe1956},
\begin{equation}
\label{eq2}
    U(0) = \int\limits_{-\infty}^{\infty}\int\limits_{-\infty}^{\infty}A(0) \,e^{j2\pi(f_x x + f_y y)} \, df_x\, df_y.
\end{equation}
Here, $f_x = \frac{\alpha}{\lambda}$ and $f_y = \frac{\beta}{\lambda}$, where $\alpha$ and $\beta$ are direction cosines of the plane wave components with respect to the $x$ and $y$ axes. Note that eq.~\eqref{eq2} is of the form of 2-D inverse Fourier transform. Hence, the field and its angular spectrum can be considered as a Fourier transform pair. A generalized form of this result can be written as
\begin{equation}
    \label{eq4}
    U(z) \xleftrightarrow{\mathcal{F}} A(z),
\end{equation}
where $U(z) \equiv U(x,y,z)$ is the field on $z = z$ plane and $A(z) \equiv A(f_x,f_y,z)$ is its angular spectrum. Suppose that the relationship between the angular spectrum at $z=0$ and $z=z$ planes is given as
\begin{equation}
    \label{eq7}
    A(z) = A(0)\,G(z)
\end{equation}
where, $G(z) \equiv G(f_x,f_y,z)$ is the propagation transfer function which characterises the propagation of the angular spectrum.

Electromagnetic waves satisfy the Helmholtz equation and therefore $U(z)$ should also satisfy it. This can be used to find an expression for $G(z)$. The Helmholtz equation is given as
\begin{equation}
    \label{eq8}
    \left(\nabla^2 + k^2\right)\,U(z) = 0,
\end{equation}
where $k = \frac{2\pi}{\lambda}$ is the wavenumber which is the magnitude of the propagation vectors $\Vec{k} = [k_x,k_y,k_z]$ of the wave components. Considering eq.~\eqref{eq4}, eq.~\eqref{eq7}, and eq.~\eqref{eq8}, it can be observed that $G(z)$ satisfies
\begin{equation}
    \label{eq9}
    \dv[2]{G(z)}{z} + k_z^2\,G(z) = 0,
\end{equation}
where
\begin{align}
    \label{eq11}
    k_z &= 2\pi\,\sqrt{\frac{1}{\lambda^2} - f_x^2 - f_y^2}.
\end{align}
An elementary solution to eq.~\eqref{eq9} can be written as,
\begin{equation}
    \label{eq12}
    G(z) = \exp \left(j2\pi z\,\sqrt{\frac{1}{\lambda^2} - f_x^2 - f_y^2}\right).
\end{equation}
Hence, eq.~\eqref{eq7} can be re-written as
\begin{equation}
    \label{eq13}
    A(z) = A(0)\,\exp \left(j2\pi z\,\sqrt{\frac{1}{\lambda^2} - f_x^2 - f_y^2}\right).
\end{equation}
This equation shows that when $f_x^2 + f_y^2 \leq \frac{1}{\lambda^2}$, the propagation of the angular spectrum along the $z$ axis introduce a different set of phase shift to each of the components of the angular spectrum~\cite{goodman2005}. However, when $f_x^2 + f_y^2 > \frac{1}{\lambda^2}$, eq.~\eqref{eq13} can be written as,
\begin{equation}
    \label{eq14}
    A(z) = A(0)\,\exp \left(-2\pi z\,\sqrt{f_x^2 + f_y^2 - \frac{1}{\lambda^2}}\right).
\end{equation}
In this case, the angular spectrum is exponentially attenuated along the $z$ axis. These wave components are known as \emph{evanescent waves} and they do not propagate energy along the $z$ axis.

\subsubsection{Implementation of the Angular Spectrum Method}

\begin{figure}[t!]
    \centering
    \includegraphics[width=0.5\linewidth]{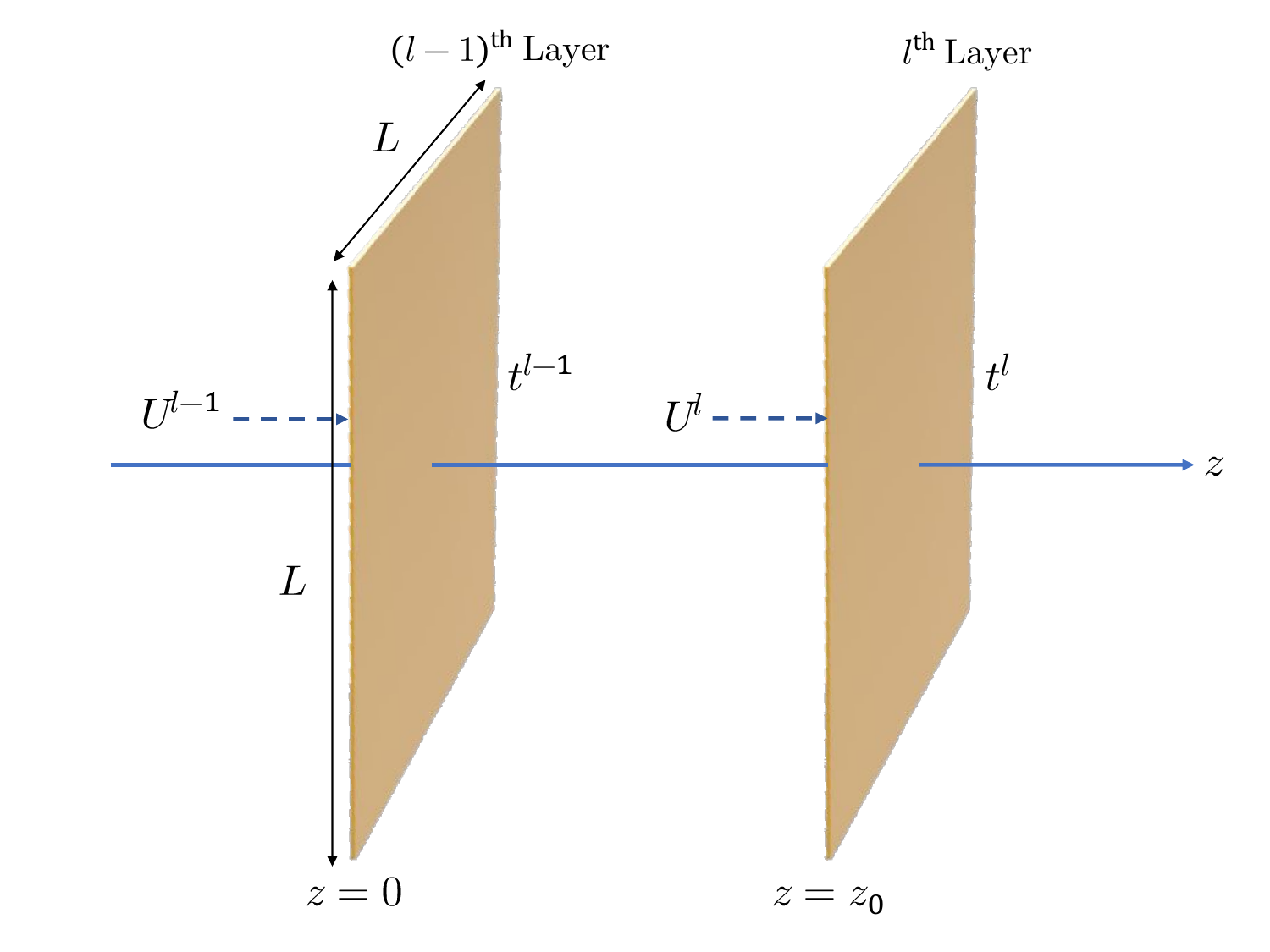}
    \caption{\textbf{Two adjacent layers of the D2NN.} The field incident on the $(l-1)^{\textrm{th}}$ layer is $U^{l-1}$ which is subjected to diffraction at the layer and results the field $U^l$ just before the next layer.}
    \label{fig:aperture_obs}
\end{figure}

The Fourier relationship between the electric/magnetic field of a light wave and its AS can be obtained using the discrete Fourier transform (DFT) which can be computed efficiently using a fast Fourier transform (FFT) algorithm. Consider two layers of a D2NN at $z = 0$ plane ($(l-1)^{\textrm{th}}$ layer) and at $z = z_0$ plane ($l^{\textrm{th}}$ layer) as shown in Fig.~\ref{fig:aperture_obs}. The input field at the $(l-1)^{\textrm{th}}$ layer is given by $U(x,y,0) \equiv U^{l-1}$ and the resulting field at the observation plane is given by $U(x,y,z_0) \equiv U^l$. The width and the height of both layers are $L$ where $t^{l-1}$ and $t^l$ are the complex transmission coefficient matrices of the two layers.

\begin{figure*}[t!]
    \centering
    \includegraphics[width=\linewidth]{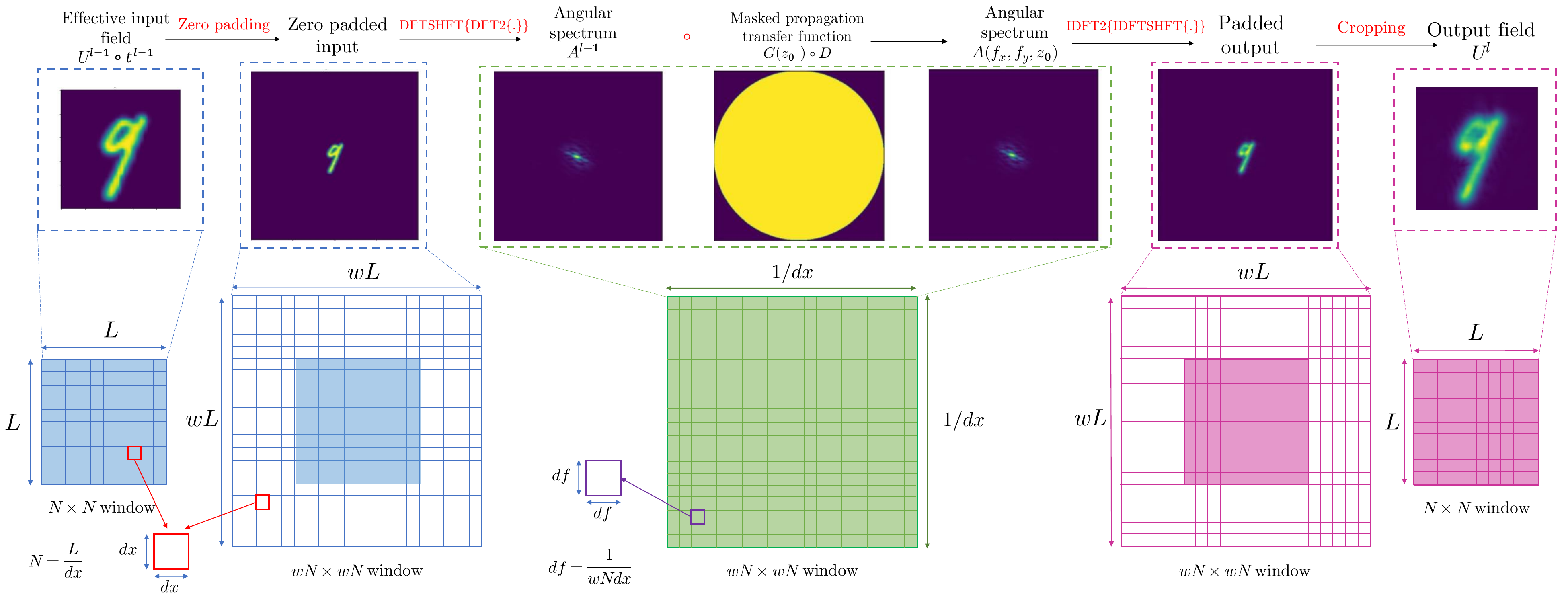}
    \caption{\textbf{Computational pipeline of the AS method.} In the figure $DFT2\{\cdot\}$ denotes the 2D-DFT operation, $DFTSHFT\{\cdot\}$ denotes the DFT shift operation, $IDFT2\{\cdot\}$ denotes the 2D-IDFT operation, $IDFTSHFT\{\cdot\}$ denotes the IDFT shift operation, and $\circ$ denotes the element-wise matrix multiplication.}
    \label{fig:fft-as_pipe}
\end{figure*}

Fig.~\ref{fig:fft-as_pipe} shows the computation pipeline of the output field using the AS method. Initially, the input field is sampled with a sampling interval of $dx$ in the spatial domain. The sampled input has a size of $N \times N$ samples, where $N = L/dx$. The sampled input is then multiplied element-wise with $t^{l-1}$ to obtain the effective field at the $(l-1)^{\textrm{th}}$ layer which is given by
\begin{equation}
    U^{(l-1)}_{\mathit{eff}} = U^{l-1} \circ t^{l-1},
\end{equation}
where $\circ$ is the element-wise matrix multiplication. 

Since the region of support of the input is bounded in the spatial domain, the region of support of its Fourier transform is unbounded and to get a more accurate AS, a higher number of samples of the Fourier transform need to be considered. Therefore, for the computation purposes a separate computational window of the size $wN \times wN$ samples is considered where $w$ is the computation window factor. The oversampling in the Fourier domain is performed by zero padding $U^{l-1}_{\mathit{eff}}$ at the boundary to match the size of the computation window. Then the 2D-DFT of the input field is taken followed by a DFT shift operation to make the center of the computation window $(0,0)$ resulting the AS given by $A^{l-1}$.

$A^{l-1}$ has spatial frequency components in the range of $\left[\frac{-f_s}{2}, \frac{f_s}{2}\right]$ in both $f_x$ and $f_y$ axes where $f_s = 1/dx$ is the sampling frequency in the spatial domain. Note that the gap between two samples in the angular spectrum is $df  = 1/wNdx$. The propagation transfer function at $z = z_0$, $G(z_0)$ is also created in the computation window with a similar number of samples as $A^{l-1}$. Since the evanescent waves do not propagate energy along the $z$-axis, they are filtered out using a mask $D$. The masked propagation transfer function is given by
\begin{equation}
    \label{eq17}
    G(z_0)\, \circ\, D = 
    \begin{dcases}
        e^{j2\pi z_0\,\sqrt{\frac{1}{\lambda^2} - f_x^2 - f_y^2}} & f_x^2 + f_y^2 \leq \frac{1}{\lambda} \\
        0 &  f_x^2 + f_y^2 > \frac{1}{\lambda}.\\
    \end{dcases}
\end{equation}
According to eq.~\eqref{eq7}, the AS of the field at $z=z_0$ is obtained by the element-wise multiplication of $A^{l-1}$ and the masked propagation transfer function. Then an inverse DFT (IDFT) shift operation is performed followed by a 2D-IDFT operation to get the padded resulting field. Finally, the padding is removed to retrieve the resulting field $U^l$.

\section{Further Experiments}
\subsection{Ablation Studies for CNN}
\label{subsec:cnn_exps}

We began experimenting with the number of linear convolutional layers for the CNN, where we increased the number of layers from 5-20 (results are shown in Table \ref{tab:cnn_layers}). Then we experimented with the kernel size and number of kernels per layer for the best performing CNNs selected from Table \ref{tab:cnn_layers} for each dataset. We also experimented with the number of kernels per layer for the 5-layer CNN for each dataset to determine if a CNN with a lesser number of layers could achieve better performance. The results of this study for each dataset are depicted in Tables \ref{tab:cnn-mnist_pi}, \ref{tab:cnn-mnist_2pi}, \ref{tab:cnn-hela_pi}, \ref{tab:cnn-hela_2pi}, \ref{tab:cnn-bacteria_2pi}.

\begin{table*}[h]
\centering
\caption{\textbf{Effect of increasing the number of layers}}
\label{tab:cnn_layers}
\begin{tabular}{lccccc}
\# layers & \multicolumn{5}{l}{\hspace{5cm} {SSIM}}\\
\cline{2-6}
 & MNIST $[0,\pi]$ & MNIST $[0,2\pi]$ & HeLa $[0,\pi]$ & HeLa $[0,2\pi]$ & Bacteria $[0,\pi]$ \\
\midrule[1pt]
5 & \textbf{0.9980} & 0.7891 & 0.9148 & 0.7895 & 0.9796\\
6 & 0.9911 & 0.7348 & 0.9114 & 0.8247 & \textbf{0.9930}\\
7 & 0.9814 & 0.7663 & 0.8995 & 0.6954 & 0.9928\\
8 & 0.9945 & \textbf{0.8010} & 0.9018 & \textbf{0.8537} & 0.9892\\
9 & 0.9847 & 0.7887 & 0.9199 & 0.7896 & 0.9874\\
10 & 0.9949 & 0.7630 & 0.9136 & 0.8063 & 0.9917\\
11 & 0.9823 & 0.7955 & 0.9295 & 0.8181 & 0.9870\\
12 & 0.9849 & 0.7549 & \textbf{0.9327} & 0.8297 & 0.9823\\
20 & 0.9867 & 0.7611 & 0.9065 & 0.6459 & 0.9854\\

\end{tabular}
\end{table*}
\FloatBarrier

\begin{table}[H]
\centering
\caption{\textbf{Hyperparameter search - MNIST $[0, \pi]$}}
\label{tab:cnn-mnist_pi}
\begin{tabular}{lccc}
\# layers & Kernel size & \# {of kernels in each layer} & SSIM \\
\midrule[1pt]
5 & 3 & 5 & {0.9984}\\
5 & 3 & 10 & 0.9974\\
5 & 5 & 1 & 0.9949\\
5 & 5 & 5 & \textbf{0.9982}\\

\end{tabular}

\end{table}
\FloatBarrier

\begin{table}[H]
\centering
\caption{\textbf{Hyperparameter search - MNIST $[0, 2\pi]$}}
\label{tab:cnn-mnist_2pi}
\begin{tabular}{lccc}
\# layers & Kernel size & \# {of kernels in each layer } & SSIM \\
\midrule[1pt]
5 & 3 & 5 & 0.7916\\
5 & 3 & 10 & 0.7756\\
8 & 5 & 1 & 0.7923\\
8 & 3 & 5 & 0.7681\\
8 & 5 & 5 & 0.7973\\

\end{tabular}

\end{table}
\FloatBarrier

\begin{table}[H]
\centering
\caption{\textbf{Hyperparameter search - HeLa $[0, \pi]$}}
\label{tab:cnn-hela_pi}
\begin{tabular}{lccc}
\# layers & Kernel size & \# {of kernels in each layer } & SSIM \\
\midrule[1pt]
5 & 3 & 5 & 0.9304\\
5 & 3 & 10 & \textbf{0.9329}\\
12 & 5 & 1 & 0.8304\\
12 & 3 & 5 & 0.9170\\
12 & 5 & 5 & 0.9297\\

\end{tabular}

\end{table}
\FloatBarrier

\begin{table}[H]
\centering
\caption{\textbf{Hyperparameter search - HeLa $[0, 2\pi]$}}
\label{tab:cnn-hela_2pi}
\begin{tabular}{lccc}
\# layers & Kernel size & \# {of kernels in each layer } & SSIM \\
\midrule[1pt]
5 & 3 & 5 & 0.8287\\
5 & 3 & 10 & 0.8051\\
8 & 5 & 1 & 0.8429\\
8 & 3 & 5 & 0.8404\\
8 & 5 & 5 & 0.7981\\

\end{tabular}

\end{table}
\FloatBarrier

\begin{table}[H]
\centering
\caption{\textbf{Hyperparameter search - Bacteria $[0, \pi]$}}
\label{tab:cnn-bacteria_2pi}
\begin{tabular}{lccc}
\# layers & Kernel size & \# {of kernels in each layer} & SSIM \\
\midrule[1pt]
5 & 3 & 5 & 0.9898\\
5 & 3 & 10 & 0.9841\\
6 & 5 & 1 & 0.9890\\
6 & 3 & 5 & 0.9876\\
6 & 5 & 5 & 0.9919\\

\end{tabular}

\end{table}


\begin{figure*}[t!]
    \centering
    \includegraphics[width=0.75\linewidth]{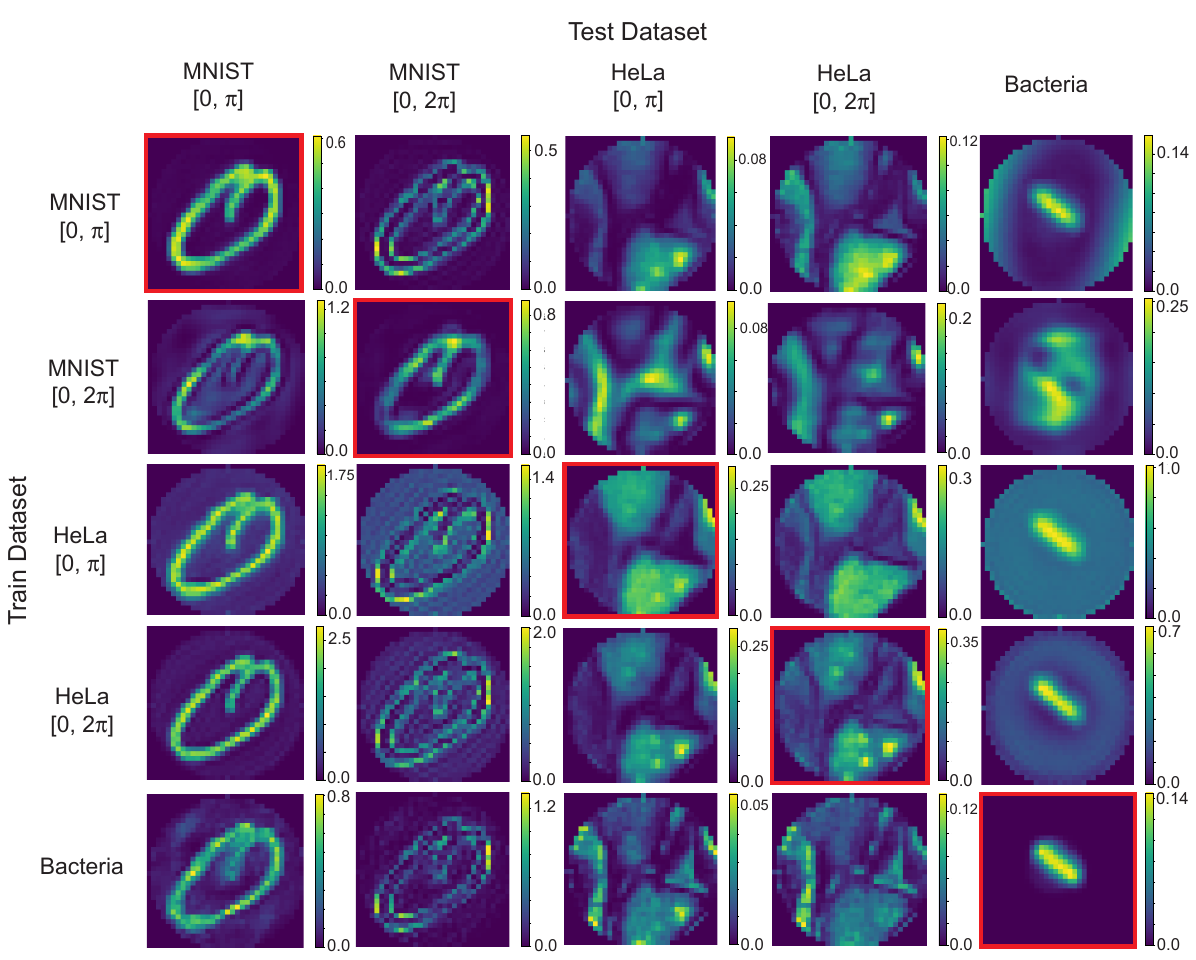}
    \caption{\textbf{Generalizability of $\phi$-LFF.} The vertical axis shows the dataset used for training and the horizontal axis shows the dataset used for testing.}
    \label{fig:generalize_lff}
\end{figure*}

\begin{figure*}[t!]
    \centering
    \includegraphics[width=\linewidth]{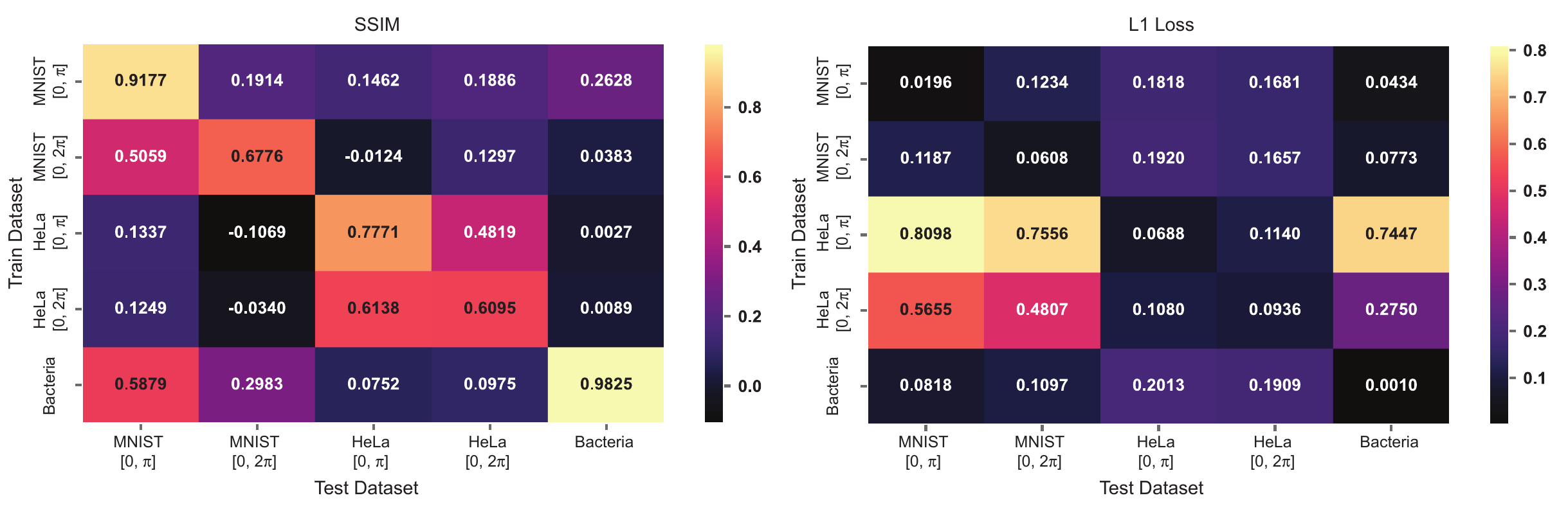}
    \caption{\textbf{Generalizability of $\phi$-LFF.} The vertical axis shows the dataset used for training and the horizontal axis shows the dataset used for testing.}
    \label{heatmap:generalize_lff}
\end{figure*}

\clearpage
\subsection{Generalizability}
\label{sec:generalizibility}

To discover how the learned optical models generalize across different datasets, we conducted a series of experiments on $\phi$-LFF, $\phi$-LRF, and $\phi$-D2NN. In particular, we evaluate the generalizability of a given optical model by training it on a particular dataset and testing it on the rest of the datasets.

Generally, we observe that the models trained on datasets with larger phase ranges perform better for datasets with similar structural features and having phase values constrained to a smaller range. But for models trained on smaller phase ranges, this generalization does not hold.


\subsubsection{$\phi$-LFF Generalizability Experiments}
\label{subsec:lff_gen}

    
    
    

The qualitative results in Figure \ref{fig:generalize_lff} and the quantitative values reflected in the heatmaps \ref{heatmap:generalize_lff} shows how generalizable the proposed $\phi$-LFF is across different datasets.

When we train the model on MNIST$[0,2\pi]$, it performs {reasonably well} on MNIST$[0,\pi]$ and performs better compared to the other three datasets. Similarly, when we train the model on HeLa$[0,2\pi]$, it also has a good SSIM score (0.6138) on HeLa$[0,\pi]$  while having better performance compared to the other three test datasets. However, training on MNIST$[0,\pi]$ and HeLa$[0,\pi]$ datasets does not guarantee {reasonable} performance on MNIST$[0,2\pi]$ and HeLa$[0,2\pi]$ datasets respectively. This demonstrates that training on the datasets with higher phase range seems to perform better on datasets with similar structural features that are spread out in a much lower phase range. This suggests that data augmentations on the phase range could help improve performance of these models.

Furthermore, we observe that training with Bacteria dataset performs {better} on MNIST$[0,\pi]$ compared to other test datasets. This also can be explained with the above argument since Bacteria and MNIST both have relatively simple structural features that are located at the center of the field of view.

\subsubsection{$\phi$-LRF Generalizability Experiments}
\label{subsec:lff_ring_gen}


    
    

\begin{figure*}[h]
    \centering
    \includegraphics[width=0.75\linewidth]{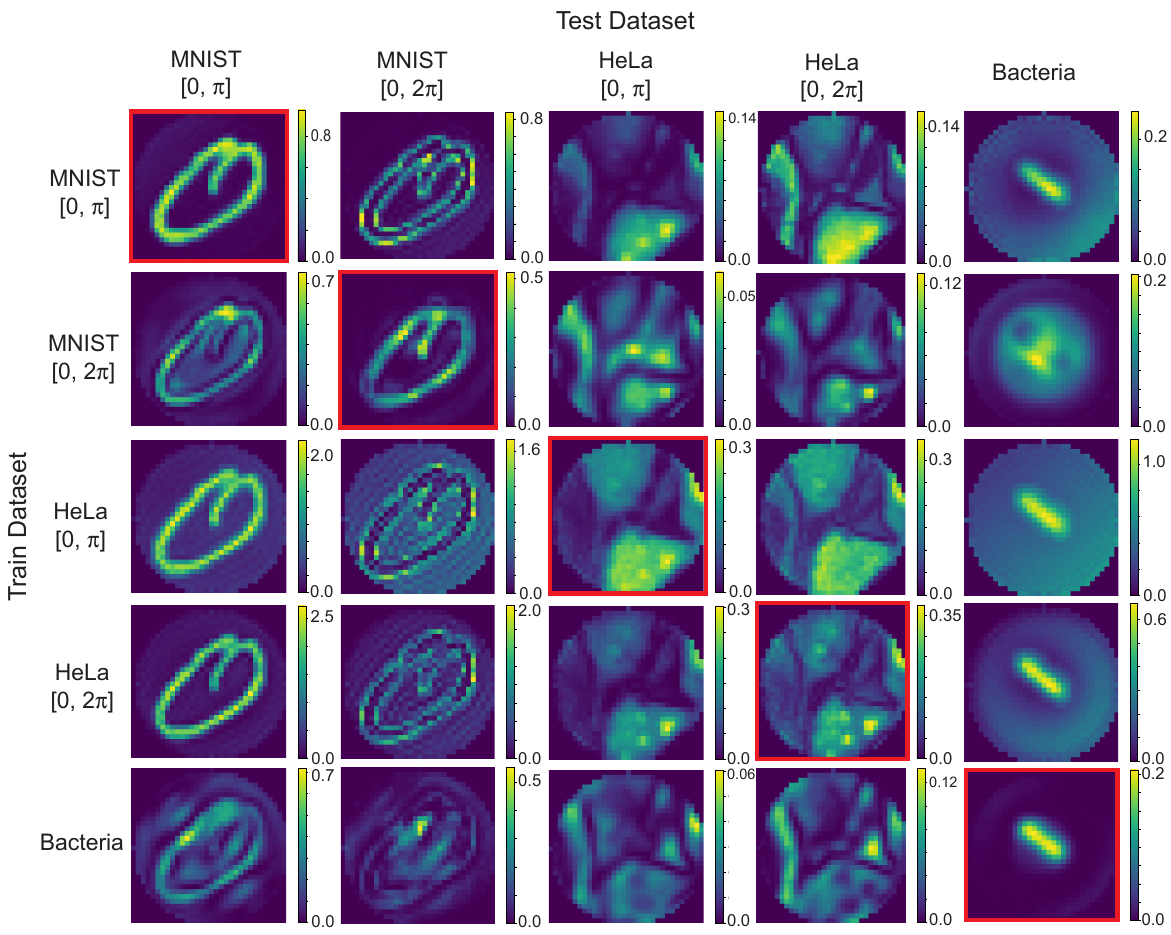}
    \caption{\textbf{Generalizability of $\phi$-LRF.} The vertical axis shows the dataset used for training and the horizontal axis shows the dataset used for testing.}
    \label{fig:generalize_lrf}
\end{figure*}

\begin{figure*}[h]
    \centering
    \includegraphics[width=\linewidth]{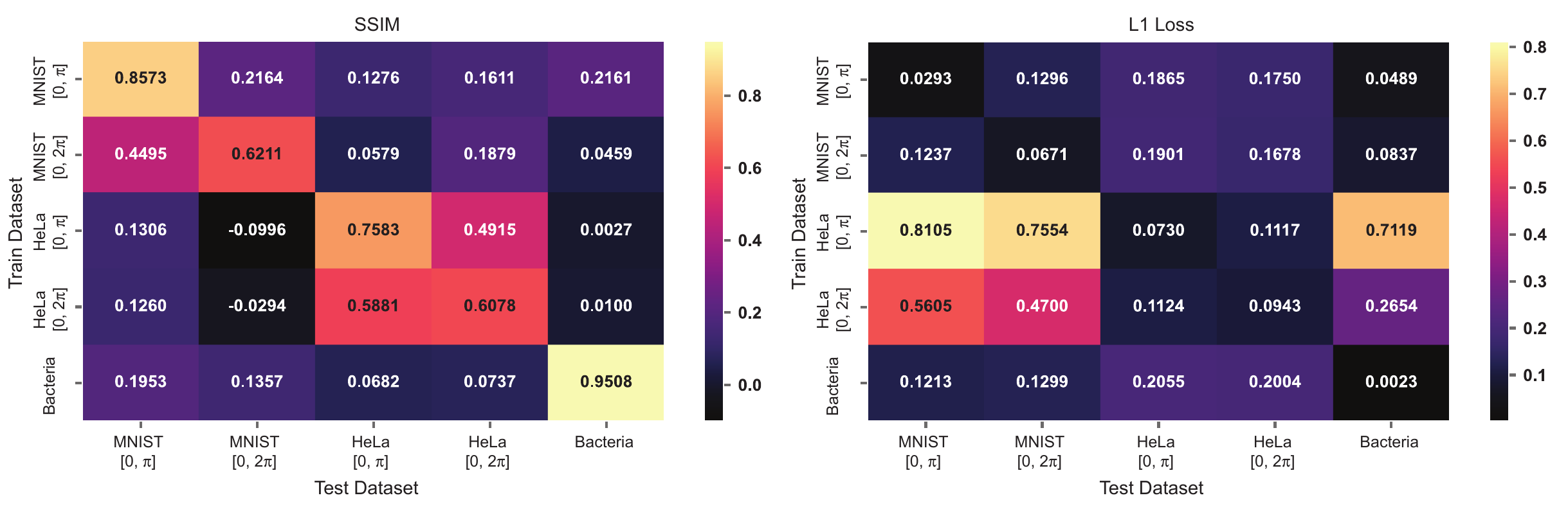}
    \caption{\textbf{Generalizability of $\phi$-LRF.} The vertical axis shows the dataset used for training and the horizontal axis shows the dataset used for testing.}
    \label{heatmap:generalize_lrf}
\end{figure*}

Similar to section \ref{subsec:lff_gen}, we conduct experiments on the models with $\phi$-LRFs. The results are shown in Fig.~\ref{fig:generalize_lrf} and heatmaps \ref{heatmap:generalize_lrf}.

Similar to previous experiments, here too, we observe similar generalization behavior. Models trained on MNIST$[0,2\pi]$ and HeLa$[0,2\pi]$ perform {better} on MNIST$[0,\pi]$ and HeLa$[0,\pi]$ compared to other test datasets. However, unlike in section \ref{subsec:lff_gen}, we do not observe the generalizability between Bacteria and MNIST$[0,\pi]$ datasets.

\subsubsection{$\phi$-D2NN Generalizability Experiments}
\label{subsec:d2nn_gen}


    
    

\begin{figure*}[h]
    \centering
    \includegraphics[width=0.75\linewidth]{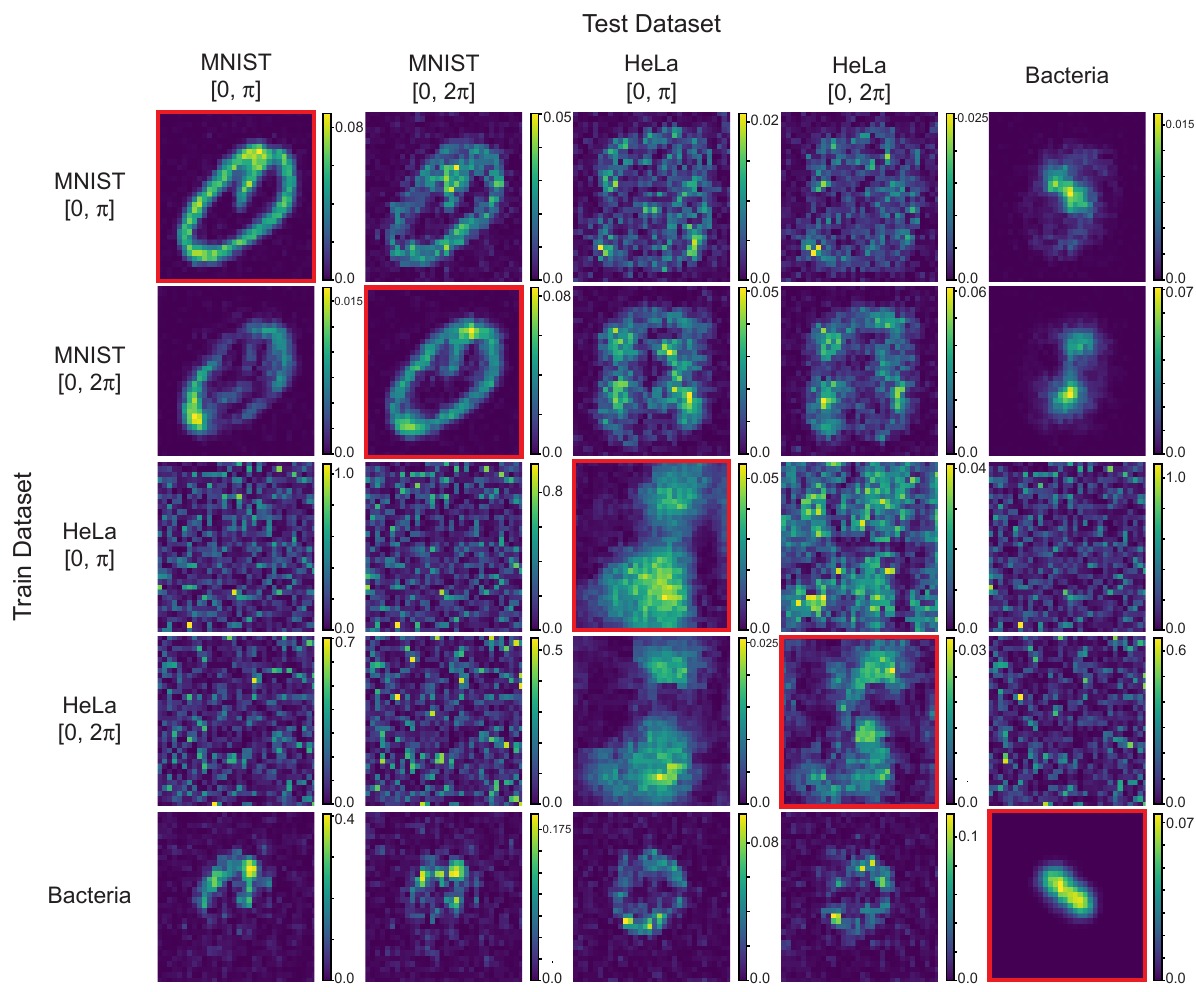}
    \caption{\textbf{Generalizability of $\phi$-D2NN.} The vertical axis shows the dataset used for training and the horizontal axis shows the dataset used for testing.}
    \label{fig:generalize_d2nn}
\end{figure*}

\begin{figure*}[h]
    \centering
    \includegraphics[width=\linewidth]{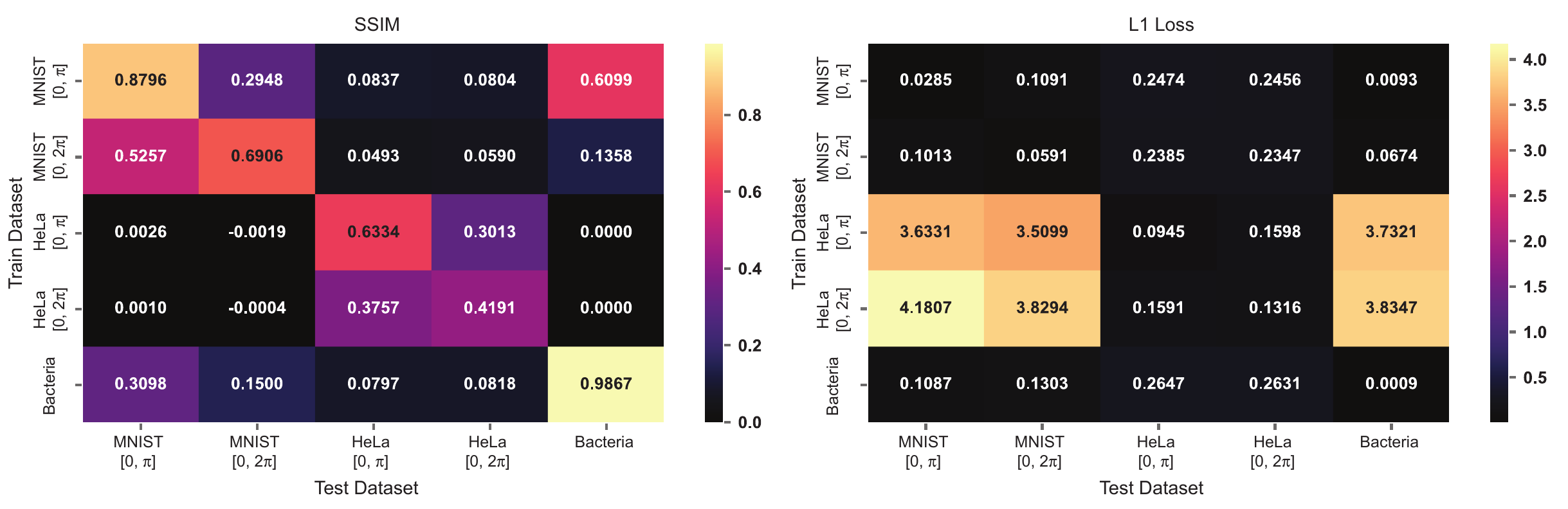}
    \caption{\textbf{Generalizability of $\phi$-D2NN.} The vertical axis shows the dataset used for training and the horizontal axis shows the dataset used for testing.}
    \label{heatmap:generalize_d2nn}
\end{figure*}

Similar to the previous sections, we conduct generalizability experiments on $\phi$-D2NN models. The results are shown in Fig. \ref{fig:generalize_d2nn} and heatmaps \ref{heatmap:generalize_d2nn}.

Similar to previous generalizability experiments, models trained on MNIST$[0,2\pi]$ and HeLa$[0,2\pi]$ perform better on MNIST$[0,\pi]$ and HeLa$[0,\pi]$ respectively. Furthermore, the $\phi$-D2NN model trained on MNIST $[0,\pi]$ performs with an SSIM score of 0.6072 on the Bacteria dataset.


\section{Mapping Design Configurations of the Trained Filter to the Experimental Setup}
\label{sec:map_simul_exp}

\begin{figure*}[t!]
    \centering
    \includegraphics[width=0.4\linewidth]{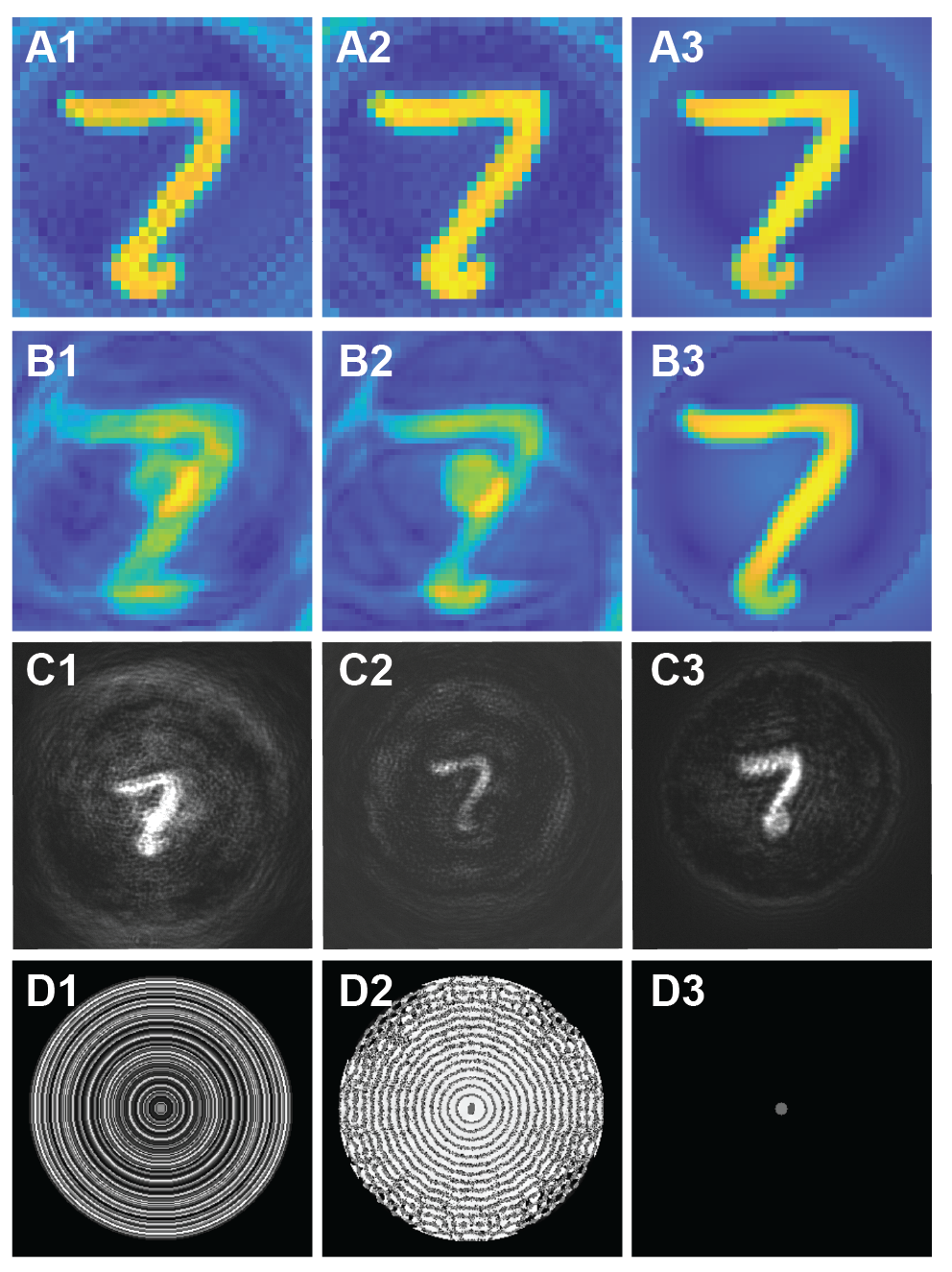}
    \caption{\textbf{Model mismatch due to up-sampling.} \textbf{(A1-3)} The numerical predictions under the training conditions (i.e., using 256x256 pixel filter) for $\phi$-LRF, $\phi$-LFF, and $\phi$-GPC filters. \textbf{(B1-3)} The corresponding numerical predictions with the up-sampled filters matching the experimental setup. \textbf{(C1-3)} The corresponding experimental results with the upsampled filters. \textbf{(D1-3)} The corresponding up-sampled filters: D1 – $\phi$-LRF, D2 – $\phi$-LFF, and D3 – $\phi$-GPC. }
    \label{fig:upsampling_mismatch}
\end{figure*}

As explained in Section IV-D our learned filters span a region of $256 \times 256$ pixels for an input phase image size of $32 \times 32$. However, in the experimental setup the input would not necessarily span within a spatial dimension of $32 \times 32$ pixels. Therefore, in the experimental setup, based on the input size, the filter implemented in the SLM would have to be resized to properly follow the simulated conditions and match the experimental dimensions for reconstruction. The effect of upsampling the learned $\phi$-LFF, $\phi$-LRF, and $\phi$-GPC filter to match the experimental setup is demonstrated in Fig.~\ref{fig:upsampling_mismatch}. The digit is clearly visible for the $\phi$-LRF (Fig.~\ref{fig:upsampling_mismatch}-C1) and $\phi$-LFF (Fig.~\ref{fig:upsampling_mismatch}-C2), however, the digits appear to have a low brightness and/or background artifacts compared to the reconstructions under the training conditions (Fig.~\ref{fig:upsampling_mismatch}-A1 and A2 respectively). This suggests that LFFs and LRFs may be sensitive to interpolation when resizing the filters. This can be alleviated by either training or finetuning the LFF/ LRF in a simulated environment that matches the experimental setup, or improving the robustness by learning a filter on an augmented dataset that captures some of the experimental setup imperfections.

\input{sec/big_table_sup}

\begin{figure*}[t!]
    \centering
    \includegraphics[width=.7\linewidth]{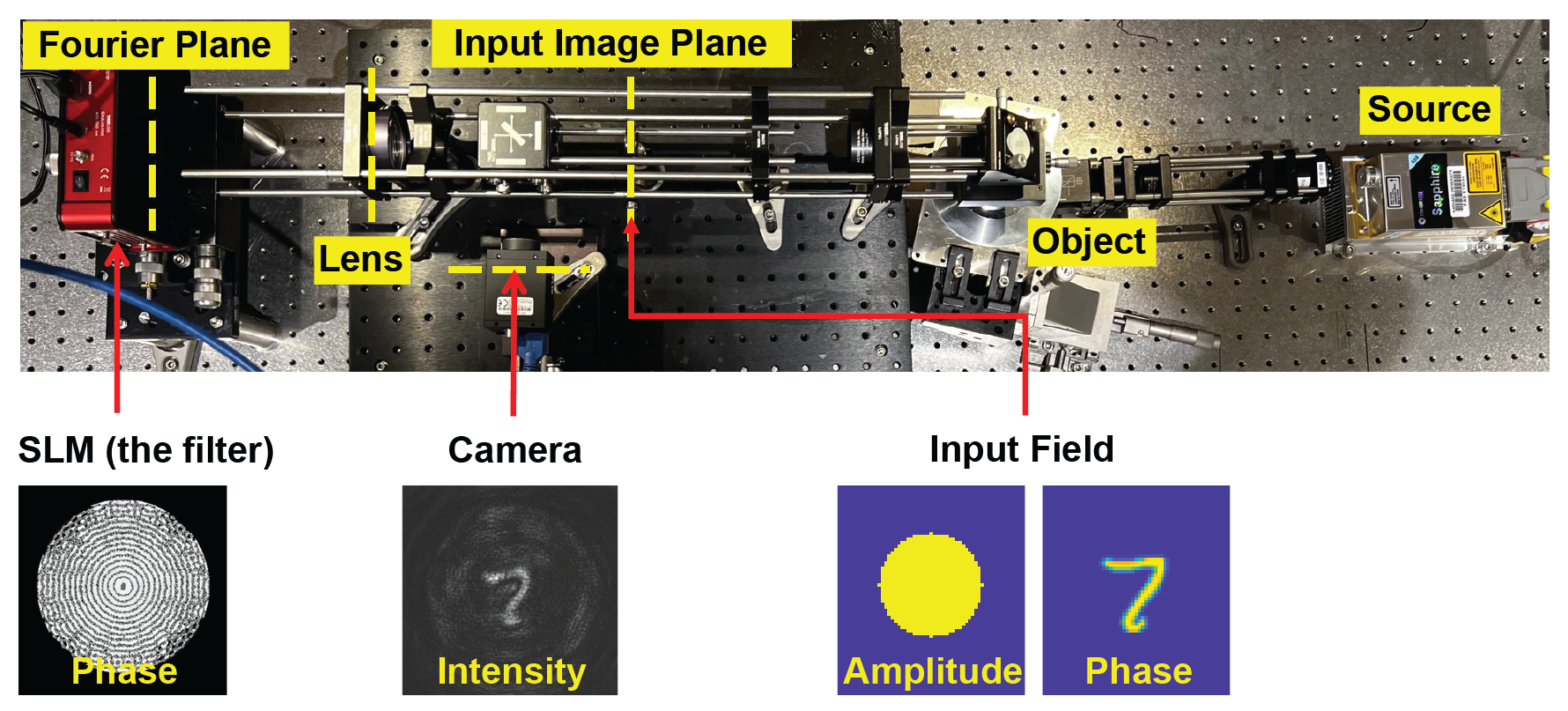}
    \caption{\textbf{A photograph of the experimental setup.}}
    \label{fig:exp_setup_photo}
\end{figure*}

{\color{black}

\section{Robustness Analysis}
\label{sec:robustness_analysis}

In this section, we provide detailed quantitative results for the robustness of trained LFF under various non-ideal conditions, including phase quantization and parameter (weight) noise. These experiments were conducted across five datasets in which those LFFs were trained: MNIST[0,$\pi$], MNIST[0,$2\pi$], HeLa[0,$\pi$], HeLa[0,$2\pi$], and Bacteria.

\subsection{Phase Quantization}
As shown in Fig. \ref{fig:robustness_plots}, the LFF demonstrates high resilience to phase quantization. The SSIM remains virtually unchanged across all datasets when quantizing the learned phase masks to 8-bit and 4-bit precisions. A significant drop in performance is only observed at 2-bit quantization (where the phase is constrained to only 4 discrete levels). In simulation, the learned designs are robust to the discrete phase resolution of commercially available spatial light modulators (SLMs), which typically provide 8-bit control.

\begin{figure*}[t!]
    \centering
    \includegraphics[width=\textwidth]{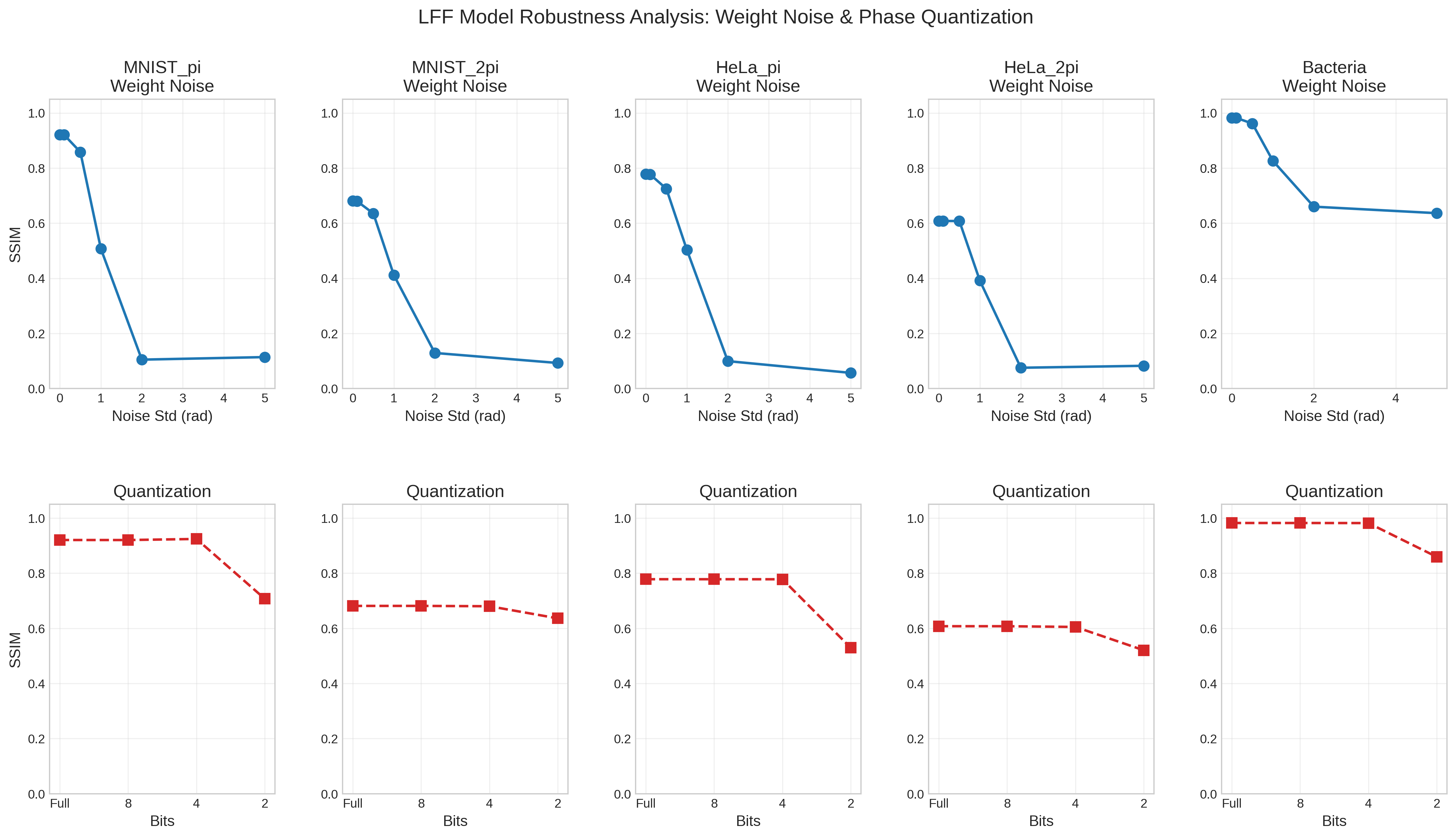}
    \caption{\textbf{Robustness of the Learnable Fourier Filter (LFF) to system non-idealities.} The top row shows the resilience of LFF models to additive Gaussian noise in the learned weights (transmission coefficients). The bottom row shows the impact of phase quantization. Performance remains stable down to 4-bit quantization and across minor weight perturbations ($\sigma < 0.1$).}
    \label{fig:robustness_plots}
\end{figure*}

\subsection{Parameter Perturbations (Weight Noise)}
The LFF also exhibits significant stability against weight perturbations. As shown in the top row of Fig. \ref{fig:robustness_plots}, the performance remains near baseline for small noise standard deviations ($\sigma \leq 0.1$ rad). Performance begins to degrade gracefully as the noise level increases, with a prominent drop occurring only as the noise standard deviation becomes a significant fraction of the learned parameter range (typically $\sigma > 0.5$ rad). This robustness to weight noise indicates that the learnable fourier filter weights are not overly sensitive to minor fabrication or calibration errors.}

%% file: sec/big_table_sup.tex
\begin{table*}[t]
\centering
\caption{\textbf{Overall quantitative results.} Performance of the Complex-valued CNN (C-CNN), GPC method, the learnable Fourier filter (LFF), and the D2NN are shown for each dataset. The best-performing cases in each performance metric for each dataset are in bold.}
\label{tab:overall_results_sup}
\resizebox{\textwidth}{!}{
\begin{tabular}{llcccccccccc}
Method & Loss & \multicolumn{2}{c}{MNIST $[0, \pi]$}  & \multicolumn{2}{c}{MNIST $[0, 2\pi]$} & \multicolumn{2}{c}{HeLa $[0, \pi]$} & \multicolumn{2}{c}{HeLa $[0, 2\pi]$} & \multicolumn{2}{c}{Bacteria $[0, \pi]$} \\ \cline{3-12} 
 & & SSIM $\uparrow$ & L1 $\downarrow$ & SSIM $\uparrow$ & L1 $\downarrow$ & SSIM $\uparrow$ & L1 $\downarrow$ & SSIM $\uparrow$ & L1 $\downarrow$  & SSIM $\uparrow$ & L1 $\downarrow$ \\ \midrule[1.5pt]   

C-CNN & $\mathcal{L}_{LT}$ & 0.9982 & 0.0041 & 0.7913 & 0.0539 & 0.9417 & 0.0248 & 0.8619 & 0.0485 & 0.9938 & 0.0008 \\  \cmidrule[0.5pt]{1-12}

\rowcolor{gray!30}
 & $\mathcal{L}_{\phi}$ & \textbf{0.9186} & 0.0201 & 0.6778 & 0.0609 & 0.7300 & 0.0914 & \textbf{0.6207} & 0.1034 & 0.9825 & \textbf{0.0007}\\
 \rowcolor{gray!30}
\multirow{-2}{*}{$\phi$-LFF} & $\mathcal{L}_{LT}$ & 0.9177 & \textbf{0.0196} & 0.6777 & 0.0608 & \textbf{0.7771} & \textbf{0.0688} & 0.6096 & \textbf{0.0937} & 0.9825 & 0.0010 \\

\multirow{2}{*}{$\phi$-LRF} & $\mathcal{L}_{\phi}$ & 0.8583 & 0.0292 & 0.6219 & 0.0671 & 0.7212 & 0.0932 & 0.6174 & 0.1039 & 0.9504 & 0.0023 \\
 & $\mathcal{L}_{LT}$ & 0.8573 & 0.0293 & 0.6211 & 0.0671 & 0.7583 & 0.0730 & 0.6078 & 0.0943 & 0.9508 & 0.0023 \\

\rowcolor{gray!30}
$\phi$-GPC$^{\star}$ & -- & 0.8466 & 0.0430 & 0.4191 & 0.0982 & 0.7297 & 0.0755 & 0.5184 & 0.1100 & 0.9479 & 0.0036 \\

\multirow{2}{*}{$\phi$-D2NN} & $\mathcal{L}_{\phi}$ & 0.8497 & 0.0337 & 0.5713 & 0.0736 & 0.4335 & 0.1371 & 0.2532 & 0.1711 & 0.9861 & 0.0009 \\
 & $\mathcal{L}_{LT}$ & 0.8796 & 0.0285 & \textbf{0.6906} & \textbf{0.0591} & 0.6334 & 0.0945 & 0.4191 & 0.1316 & \textbf{0.9867} & 0.0009 \\  \cmidrule[0.5pt]{1-12}

\rowcolor{gray!30}
 & $\mathcal{L}_{\phi}$ & 0.9225 & 0.0202 & 0.6801 & 0.0608 & 0.7301 & 0.0914 & \textbf{0.6202} & 0.1033 & 0.9818 & 0.0010 \\
\rowcolor{gray!30}
\multirow{-2}{*}{LFF} & $\mathcal{L}_{LT}$ & 0.9205 & \textbf{0.0196} & 0.6814 & 0.0608 & 0.7783 & \textbf{0.0688} & 0.6078 & \textbf{0.0937} & 0.9823 & 0.0010 \\

GPC$^{\star}$ & -- & 0.9036 & 0.0320 & 0.4406 & 0.0974 & \textbf{0.7786} & 0.0748 & 0.5509 & 0.1091 & 0.9600 & 0.0029 \\

\rowcolor{gray!30}
 & $\mathcal{L}_{\phi}$ & 0.8325 & 0.0357 & 0.5350 & 0.0776 & 0.4548 & 0.1302 & 0.5592 & 0.1321 & 0.9901 & 0.0007 \\
\rowcolor{gray!30}
\multirow{-2}{*}{D2NN} & $\mathcal{L}_{LT}$ & \textbf{0.9433} & 0.0201 & \textbf{0.7703} & \textbf{0.0486} & 0.6655 & 0.0936 & 0.4942 & 0.1274 & \textbf{0.9926} & \textbf{0.0007} \\

 \midrule[1.5pt]

\multicolumn{7}{l}{$^{\star}$ \footnotesize Output is scaled for each dataset (see supplementary section \ref{subsec:GPC_region_select})}

\end{tabular}}

\end{table*}